\documentclass[plaindraft]{jpp-AAS}
\usepackage{graphicx,natbib}
\usepackage{epstopdf, epsfig}
\usepackage{amsfonts,amsmath,amssymb}
\usepackage[pdftex,colorlinks,citecolor=blue]{hyperref}
\usepackage{bm}
\usepackage{xcolor}
\newcommand{\mockalph}[1]{}

\ifCUPmtlplainloaded \else
  \checkfont{eurm10}
  \iffontfound
    \IfFileExists{upmath.sty}
      {\typeout{^^JFound AMS Euler Roman fonts on the system,
                   using the 'upmath' package.^^J}%
       \usepackage{upmath}}
      {\typeout{^^JFound AMS Euler Roman fonts on the system, but you
                   dont seem to have the}%
       \typeout{'upmath' package installed. JPP.cls can take advantage
                 of these fonts, if you use 'upmath' package.^^J}%
       \providecommand\upi{\upi}%
      }
  \else
    \providecommand\upi{\upi}%
  \fi
\fi

\ifCUPmtlplainloaded \else
  \checkfont{msam10}
  \iffontfound
    \IfFileExists{amssymb.sty}
      {\typeout{^^JFound AMS Symbol fonts on the system, using the
                'amssymb' package.^^J}%
       \usepackage{amssymb}%
       \let\le=\leqslant  
         
      }{}
  \fi
\fi

\ifCUPmtlplainloaded \else
  \IfFileExists{amsbsy.sty}
    {\typeout{^^JFound the 'amsbsy' package on the system, using it.^^J}%
     \usepackage{amsbsy}}
    {\providecommand\boldsymbol[1]{\mbox{\boldmath $##1$}}}
\fi

\newcommand{\bh}{\hat{\bm{b}}}

\newcommand{\iddt}{{\partial}/{\partial t}}
\newcommand{\DDt}{\frac{D}{D t}}

\newcommand{\iDDt}{{D}/{D t}}

\newcommand{\Dp}{\Delta p}
\newcommand{\dbint}{\delta b_{\mathrm{int}}}
\newcommand{\dbturb}{\delta b_{\mathrm{turb}}}
\newcommand{\reyint}{\mathrm{It}_{\mathrm{Brag}}}

\newcommand{\simc}{\!\sim\!}
\newcommand\bb[1]{\mbox{\boldmath{$#1$}}}
\renewcommand{\bcdot}{\,\bb{\cdot}\,}
\newcommand{\bdbldot}{\,\bb{:}\,}
\newcommand{\grad}{\bb{\nabla}}
\newcommand{\bbgu}{\bh\bh\bdbldot\grad\bb{u}}
\newcommand{\omA}{\omega_{\rm A}}
\newcommand{\valf}{v_{\rm A}}
\newcommand{\const}{{\rm const}}
\newcommand{\visbrag}{\mu_{{\rm Brag}}}

\title[Magneto-immutable plasma turbulence]{Magneto-immutable turbulence in weakly collisional plasmas}

\author[J.~Squire and others]%
{J.~Squire$^{1,2}$
 \thanks{Email address for correspondence: jonathan.squire@otago.ac.nz},
 A.~A.~Schekochihin$^{3,4}$, E.~Quataert$^{5}$ 
and M.~W.~Kunz$^{6,7}$}

\affiliation{$^1$Physics Department, University of Otago, 730 Cumberland St., Dunedin 9016, New Zealand\\[\affilskip]
$^2$TAPIR, Mailcode 350-17, California Institute of Technology, Pasadena, CA 91125, USA\\[\affilskip]
$^3$The Rudolf Peierls Centre for Theoretical Physics, University of Oxford, Clarendon Laboratory, Parks Road, Oxford, OX1 3P4, UK\\[\affilskip]
$^4$Merton College, Oxford OX1 4JD, UK\\[\affilskip]
$^5$Astronomy Department and Theoretical Astrophysics Center, University of California, Berkeley, CA 94720, USA\\[\affilskip]
$^6$Department of Astrophysical Sciences, Princeton University, Peyton Hall, Princeton, NJ 08544, USA\\[\affilskip]
$^7$ Princeton Plasma Physics Laboratory, PO Box 451, Princeton, NJ 08543, USA}

\pubyear{2018}
\volume{}
\pagerange{}
\date{?; revised ?; accepted ?. - To be entered by editorial office}

\begin{document}
\maketitle



\begin{abstract}
We propose that pressure anisotropy causes weakly collisional turbulent plasmas to self-organize so as to resist changes in magnetic-field strength. We term this effect ``magneto-immutability'' by analogy with incompressibility (resistance to changes in pressure). The effect is important when the pressure anisotropy becomes comparable to the magnetic pressure, suggesting that in collisionless, weakly magnetized (high-$\beta$) plasmas its dynamical relevance is similar to that of incompressibility. Simulations of magnetized turbulence using the weakly collisional Braginskii model show that magneto-immutable turbulence is surprisingly similar, in most statistical measures, to critically balanced MHD turbulence. However, in order to minimize magnetic-field variation, the flow direction becomes more constrained than in MHD, and the turbulence is more strongly dominated by magnetic energy (a nonzero ``residual energy'').  These effects represent key differences between  pressure-anisotropic and fluid turbulence, and should be observable in the $\beta\gtrsim1$ turbulent solar wind.
\end{abstract}


\section{Introduction}
Many magnetized astrophysical plasmas -- for example, the solar wind
 and the intracluster medium of galaxy clusters -- are turbulent and  
 weakly collisional, with particle mean free paths that are comparable to, or exceed, the scales of plasma motions. Despite this scale hierarchy, it is broadly assumed that such plasmas can be described by single-fluid magnetohydrodynamics (MHD), at least on scales much larger than the plasma's kinetic microscales (e.g., the ion gyroradius $\rho_i$ or skin depth). Indeed, there are certain situations in which this simplification can be justified rigorously \citep[e.g.,][]{Kulsrud:1980tm,Schekochihin:2009eu}. 
In this work, we show that there exists a significant dynamical effect in weakly collisional plasmas that is \emph{not} captured 
by the MHD model. It affects plasmas whose thermal energies are comparable to their magnetic 
energy, $\beta\equiv8\pi p_{0}/B^{2}\gtrsim1$ (where $p_{0}$ is the thermal pressure and $B=|\bm{B}|$ is the magnetic field strength). This effect, which we call ``magneto-immutability,'' is the tendency of the plasma motions to self-organize so as to resist changes in magnetic-field strength.

Magneto-immutability arises from the dynamical effects of \emph{pressure anisotropy},
\begin{equation}
\Dp\equiv p_\perp - p_\parallel,\label{eq:Dp}
\end{equation}
 which is  the
difference between the thermal pressures perpendicular ($\perp$) and parallel ($\parallel$) to the magnetic field.
Pressure anisotropy is generated locally whenever and wherever
$B$ changes slowly 
in a plasma with the ion collision frequency $\nu_{c}$  much smaller than the gyrofrequency $\Omega_{i}$ (\citealp{CGL:1956}; while
the same is true for electrons, ion microphysical parameters are most relevant for the effects studied here). 
Although pressure anisotropy is well studied in solar-wind plasmas \citep{2002GeoRL..29.1839K,Bale:2009de}, 
most authors have focused on microscale kinetic instabilities that are excited if $|\Dp|$ becomes too large, rather than on the 
dynamical feedback of $\Dp$ on the large-scale motions (but see \citealp{Squire:2016ev,Squire:2017ej,2016JPlPh..82f9001H,2017PhPl...24g2306Y}). The latter is the focus of this work.

The dynamical effects of pressure anisotropy that lead to magneto-immutability are best described by analogy with the more familiar concept of 
incompressibility. Just as density fluctuations
are minimized by the pressure force ($-\grad p$) because it drives flows away from compressions, magnetic-field-strength fluctuations are minimized by the pressure-anisotropy force $\grad\bcdot(\bh\bh\,\Dp)$,
which drives field-aligned flows towards or away from large-magnitude ``magneto-dilations,'' i.e., fluctuations for which $\bbgu\equiv\bh\bcdot(\bh\bcdot\grad\bm{u})\neq{0}$ (where 
$\bh$ is the
unit vector in the direction of the magnetic field).
A flow becomes incompressible when the time scales associated with compressive motions are short 
compared to other motions of the plasma. Likewise, 
a flow is magneto-immutable when dynamically large pressure anisotropies develop quickly compared to other important time scales (e.g., the Alfv\'en period). 
It is widely appreciated in plasma physics that weakly collisional  plasmas cannot support motions that involve a linear perturbation to $B$ (e.g., slow waves), either due to viscous or collisionless  damping \citep{Barnes1966}. Our contribution in this work is to suggest that such ideas apply equally well to nonlinear motions in a turbulent environment, \emph{viz.,} that the resistance to changes in $B$ operates as a general self-organization principle for kinetic plasmas. 



Magneto-immutability can be important whenever  $\Dp$ generated by plasma motions
approaches $B^{2}$. In this article, we focus on its relevance to Alfv\'enic turbulence, which is important in a wide 
range of space and astrophysical plasmas. Magneto-immutability
 occurs for turbulence amplitudes $\delta B_{\perp}/B$ approaching the ``interruption limit'' (see \S\ref{sec:sub:interruption.description} below), 
above which linearly polarized shear Alfv\'en waves do not propagate \citep{Squire:2016ev,Squire:2016ev2}. 
This implies that weakly collisional plasmas, our focus in this work, are approximately magneto-immutable for $\beta\gtrsim\nu_{c}/\omega>1$ (for trans-Alfv\'enic motions with $\delta B_{\perp}\simc B$), where $\omega$ is the characteristic frequency of the motion.
In contrast, for collisionless plasmas such as the solar wind, magneto-immutability likely plays a role in turbulent self-organization for $\beta$ approaching or exceeding ${\sim}1$ (for trans-Alfv\'enic turbulence), and
should be of similar dynamical importance to incompressibility.
Intriguingly, a variety of {\it in situ} observations of the turbulent solar wind
have found that the magnetic field  preferentially oscillates in such a way that  $B$ remains nearly constant  \citep{Lichtenstein1980,Tsurutani1994,2001P&SS...49.1201B}, a phenomenon often referred to as ``spherical polarization'' \citep{Vasquez1998}. While these observations  provide suggestive evidence 
that our theory may be relevant in the collisionless solar wind, other explanations for spherical polarization do exist 
(e.g., \citealt{Barnes1974,2008JGRA..113.8110B,Tenerani2018}) and further work is needed to make 
more detailed falsifiable predictions in the collisionless regime.

Following a brief review of the physics of shear-Alfv\'en-wave interruption in \S\ref{sec:sub:interruption.description}, the remainder of this paper has two main parts. First, in \S\ref{sec:general.ramblings},  we argue heuristically for the importance of 
magneto-immutability, relying heavily 
on  parallels between pressure anisotropy and compressional motions. Second, in \S\ref{sec:sims}, we present a set of 
Alfv\'enic-turbulence simulations using  the weakly collisional Braginskii MHD model, 
 the simplest model that contains the necessary physics. 
These two parts  are interdependent: 
the simulations  validate some of the key ideas and assumptions used in the physical discussion, also showing
the ways in which magneto-immutable turbulence is nonetheless similar to standard Alfv\'enic turbulence. The arguments in \S\ref{sec:general.ramblings} suggest that magneto-immutability applies more generally to weakly collisional turbulence, 
not being limited to the regime of validity of the specific model ({\em viz.}, Braginskii MHD) employed in our simulations.

\begin{figure}
\begin{center}
\includegraphics[width=0.45\columnwidth]{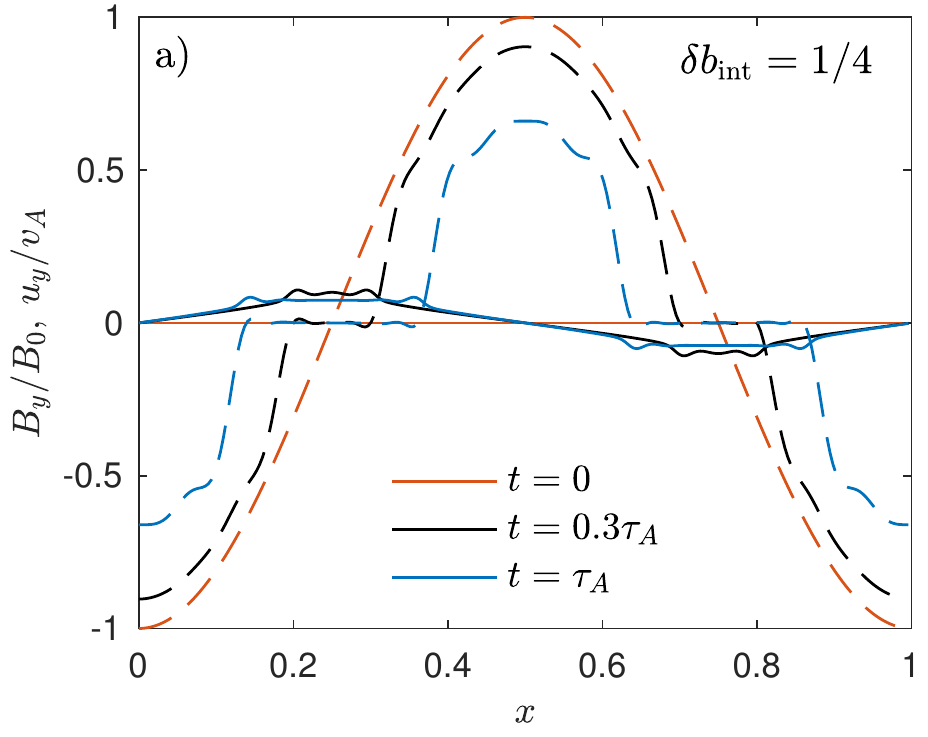}~\includegraphics[width=0.45\columnwidth]{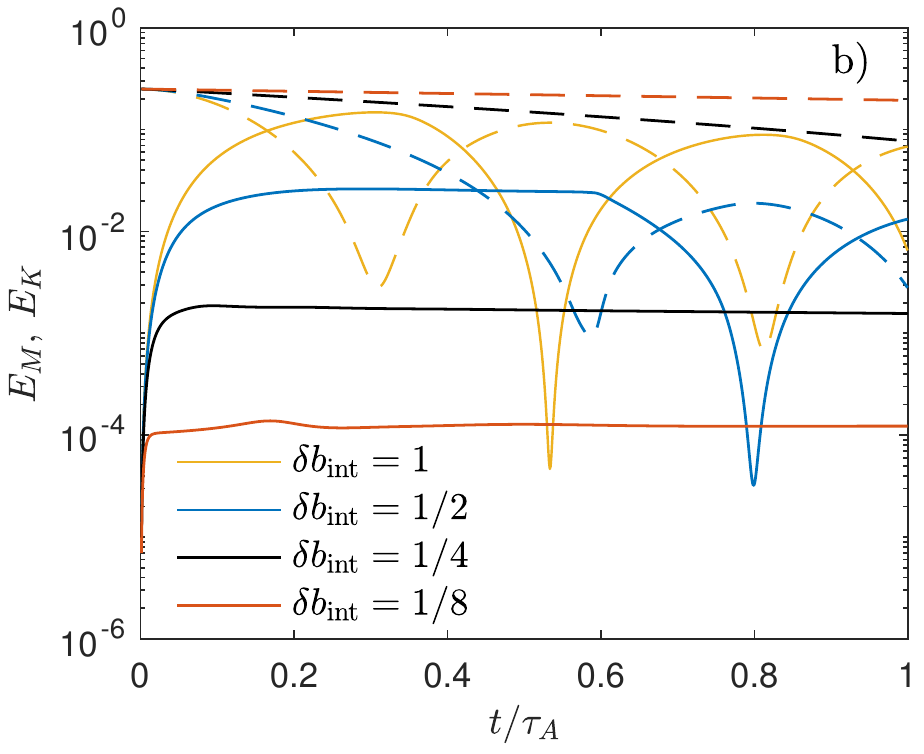}
\caption{Interruption  of  linearly polarized shear Alfv\'en waves in the Braginskii MHD model, which is used for the turbulence 
simulations presented in \S~\ref{sec:sims}. Here we show results from simulations in one dimension, starting from a perpendicular magnetic perturbation $\delta B_{\perp}/B_{0}=1$ [$B_{y}=-B_{0}\cos(2\pi x)$] with background magnetic field $\bm{B}_{0}=B_{0}\hat{\bm{x}}$. In standard MHD, these initial conditions lead to standing-wave oscillations with period $\tau_{A}$.  (a) Snapshots of the perpendicular 
velocity $u_{y}(x)/\valf$ (solid lines) and perpendicular magnetic field  $B_{y}(x)/B_{0}$ (dashed lines) for $\dbint=1/4$ (Braginskii viscosity $\visbrag\equiv\nu_{c}^{-1}p_{0}=6$; see \S\ref{sec:sub:alfvenic}). We show snapshots at $t=0$ (red lines), $t=0.3\tau_{A}$ (black lines), and $t=\tau_{A}$ (blue lines).   (b) Time evolution of kinetic energy ($E_{K}$, solid lines) and magnetic energy ($E_{M}$, dashed lines) 
 at different $\dbint$ as labeled, from $\dbint=1$ to $\dbint=1/8$ (the black curves show the full time evolution of the wave in 
panel (a)). When the Braginskii viscosity is sufficiently large so that $\delta B_{\perp}/B_{0}\gtrsim \dbint$, the system  no longer supports shear Alfv\'enic oscillations; perturbations simply decay with $E_{M}\gg E_{K}$  until they can oscillate freely at amplitudes below $\dbint$. }
\label{fig:1D.interruption}
\end{center}
\end{figure}

\subsection{Interruption of Alfv\'enic perturbations}\label{sec:sub:interruption.description}
A common concept discussed throughout this work is that of ``interruption'' of Alfv\'enic fluctuations, first introduced in \citet{Squire:2016ev}. It is helpful to  review briefly the physics of  interruption  here, 
both for  the convenience of the reader and in order to highlight the surprising nature of some of our findings. Interruption 
is a nonlinear effect that occurs when the change in the magnetic-field strength in an oscillating, linearly polarized 
shear Alfv\'en wave is sufficiently large to cause the pressure anisotropy to reach the parallel firehose threshold, $\Dp=-B^{2}/4\pi$. 
This is achieved for wave amplitudes $\delta B_{\perp}/B_{0}$ exceeding the ``interruption limit''
 \begin{equation}
\dbint\equiv
\begin{cases}\,2\beta^{-1/2},&\nu_{c}<\omA\:\text{(collisionless),}\\ 
\,2\beta^{-1/2}\left(\dfrac{\nu_{c}}{\omA}\right)^{1/2},& \nu_{c}>\omA\:\text{(weakly collisional), }
\end{cases}\label{eq:interruption.limit}
\end{equation}
 where $\omA = k_{\parallel}\valf$
 is the Alfv\'en frequency.
 The limit is particularly relevant because if $\Dp$ reaches the firehose threshold, then the magnetic tension, which is the restoring force for shear Alfv\'en waves, is nullified. The wave thus stops oscillating -- i.e., it is ``interrupted.'' This implies
 that plasmas cannot support linearly polarized shear Alfv\'en waves above the amplitude \eqref{eq:interruption.limit}. 
 Although the detailed dynamics of interrupted waves (i.e., fluctuations with $\delta B_{\perp}/B_{0}\gtrsim \dbint$) differ between the collisionless and weakly collisional regimes \citep{Squire:2016ev2} and depend on microinstabilities \citep{Squire:2017ej}, the waves always  become 
 strongly magnetically dominated, with $\langle B^{2}\rangle\gg\langle u^{2}\rangle$ and $\delta u_{\perp}/\valf\lesssim \dbint$. In the weakly collisional 
 regime, the focus of our study here, the magnetic field of an interrupted shear Alfv\'en wave decays  to below the interruption limit \eqref{eq:interruption.limit} over the timescale $t_{\mathrm{decay}}\sim \delta b_{0}^{2}\,\beta/\nu_{c}$, while the velocity perturbation remains very small (here $\delta b_{0}$ is the initial magnetic perturbation amplitude; see \citealt{Squire:2016ev2}). 
 In figure \ref{fig:1D.interruption}, we show some examples of wave interruption in the weakly collisional Braginskii MHD model (see \S\ref{sec:sub:alfvenic}) at parameters
 chosen to match those of the turbulence simulations presented in \S\ref{sec:sims} ($\dbint$ from $1/8$ to $1$, with initial perturbation amplitudes $\delta B_{\perp}/B_{0}=1$). 
Note that for propagating or standing circularly polarized shear Alfv\'en
 waves, the magnetic field remains constant in time, so the interruption limit does not apply.

 Our study here is designed to examine the influence of wave interruption on Alfv\'enic turbulence. The 
  now-standard ``critical balance'' paradigm \citep{Goldreich:1995hq,Goldreich1997} 
  posits that linear (shear-Alfv\'en-wave) and nonlinear time scales are comparable 
  at all spatial scales in MHD turbulence. An immediate corollary is that if wave 
  time scales are significantly modified due to wave interruption (which can occur at  
  low amplitudes for $\beta\gg1$), then the turbulent cascade should also be strongly modified.
 Further, in the weakly collisional regime,  $\dbint\propto \omA^{-1/2}\propto k_{\parallel}^{-1/2}$ has the same scaling as critically balanced fluctuations ($\delta u_{\perp}\propto k_{\parallel}^{-1/2}$), 
 suggesting that interruption effects should be important at all scales if they are important at the outer scale. 
  Alternatively, one could state that, for outer-scale fluctuation amplitudes $\delta u_{\perp}/\valf\sim\delta B_{\perp}/B_{0}\gtrsim\dbint$, pressure anisotropy is expected to be a stronger nonlinearity than the usual MHD nonlinearities
  across all scales of the turbulent  cascade.
This nonlinearity inhibits the oscillation of Alfv\'enic fluctuations (see figure~\ref{fig:1D.interruption}), which seems to suggest  that turbulence may not be possible for fluctuation amplitudes that exceed 
  the interruption limit. This prediction is borne out in one dimension: stochastically driving  linearly polarized shear Alfv\'en waves, one finds that the amplitude of velocity fluctuations is limited by \eqref{eq:interruption.limit}. Likewise, in figure~\ref{fig:1D.interruption}, we see that the 
  kinetic energy of decaying shear Alfv\'enic perturbations is very small for $\dbint\lesssim 1/2$. However, 
  we will show in what follows that  three-dimensional turbulence changes its characteristics 
  to avoid this scenario, becoming ``magneto-immutable'', while still supporting a turbulent cascade. 
  %
  %

%
%
%
\section{Magneto-immutable Alfv\'enic turbulence}\label{sec:general.ramblings}

Our starting point is the set of MHD equations with a pressure-anisotropy stress in the momentum equation:
\begin{align}
\rho \DDt\bm{u} &= -\grad \left( p_{\perp} + \frac{B^{2}}{8\pi} \right) + \frac{1}{4\pi}\bm{B}\bcdot\grad\bm{B}+\grad\bcdot \bigl(\bh\bh\,\Dp\bigr),\label{eq:MHD.u}\\
\DDt \bm{B} &= \bm{B}\bcdot\grad\bm{u} - \bm{B}\,\grad\bcdot \bm{u},\label{eq:MHD.b}\\
\DDt\Dp &=\,(p_{\perp}+2p_{\parallel})\,\bbgu+(p_{\parallel}-2p_{\perp})\grad\bcdot\bm{u} \nonumber\\
\mbox{} &\quad +\mathcal{Q}(q_{\perp},q_{\parallel}) - 3\nu_{c}\Dp.\label{eq:MHD.Dp}
\end{align}
Here $\rho$ is the mass density, $\bm{u}$ is the flow velocity, $\bm{B}$ is the magnetic field, 
$\iDDt\equiv\iddt+\bm{u}\bcdot\grad$ is the convective derivative, 
and $\mathcal{Q}(q_{\perp},q_{\parallel})$ parameterizes the effects of heat fluxes (see, e.g., \citealt{CGL:1956,Snyder:1997fs,2015JPlPh..81a3203S,Squire:2016ev2} for explicit reference to the equations for $p_{\perp}$ and $p_{\|}$ individually, and for discussion of $\mathcal{Q}(q_{\perp},q_{\parallel})$). The Alfv\'{e}n speed is $\valf\equiv\,B/\sqrt{4\pi\rho}$. Throughout this work, we consider only subsonic dynamics with $\grad\bcdot\bm{u}\approx0$. Equations~\eqref{eq:MHD.u}--\eqref{eq:MHD.Dp} may be derived directly from the kinetic equations \citep{Kulsrud:1980tm,Schekochihin:2010bv} by assuming collisional (or cold) electrons and using the gyrotropy of the ion distribution on scales much larger 
than the gyroradius. 
They provide the simplest well-justified model for plasma dynamics on scales much larger than $\rho_{i}$. 

\subsection{Magneto-immutability and incompressibility}\label{sec:sub:basic.description}

Although a complete solution to \eqref{eq:MHD.u}--\eqref{eq:MHD.Dp} requires specifying $\mathcal{Q}(q_{\perp},q_{\parallel})$ with a kinetic solution or closure, 
let us proceed for the moment without doing so. We draw analogies between the 
pressure-anisotropy force and the more familiar $\grad p$ force.  In all 
fluid-like equations of state, pressure is coupled to flow divergences: it increases in compressions ($\grad\bcdot\bm{u}=\delta_{rs}\nabla_{s}u_{r}=-D\ln\rho/Dt<0$) and decreases 
in rarefactions ($\grad\bcdot\bm{u}>0$). The pressure force ($-\grad p$)  isotropically drives the flow away from regions of large $p$, thus pushing fluid away from compressions and towards rarefactions. 
 This naturally leads to incompressibility, when pressure forces dominate over others in the system, rapidly eliminating compressional motions.

Similar ideas apply to pressure anisotropy and magneto-immutability.
From \eqref{eq:MHD.Dp}, we see that pressure anisotropy is driven 
by ``magneto-dilations,'' where $\bbgu=\hat{b}_{s}\hat{b}_{r}\nabla_{s}u_{r}=D\ln{B}/Dt+\grad\bcdot\bm{u}\neq0$.
The pressure-anisotropy stress in \eqref{eq:MHD.u} has the form 
$\grad\bcdot(\bh\bh\,\Dp)=\nabla_{r}(\hat{b}_{i}\hat{b}_{r}\Dp)$, and is akin to an anisotropic version of $-\grad p=-\nabla_{r}(\delta_{ir}p)$: it 
is a force that acts in a direction nearly aligned with $\bh$ (so long as $\bh$ does not vary significantly in space),
and arises due to variations in $\Dp$ along the $\bh$ direction.\footnote{For example, if the field is straight $\bh=\hat{\bm{x}}$, then $\grad\bcdot(\bh\bh\,\Dp)=(\partial_{x}\Dp)\hat{\bm{x}}$.}
We thus expect that the pressure-anisotropy stress will drive field-aligned flows that  minimize $\bbgu\approx D\ln B/Dt$.
Such a flow will resist changes in the magnetic-field strength; i.e., it will approach ``magneto-immutablity.''

Note that there is no requirement that incompressibility and magneto-immutability act separately. 
Indeed, for trans-Alfv\'enic  ($\delta B_{\perp}\simc B$) turbulence in a  collisionless plasma, both effects 
can be of the same order. In this case, it will be important to consider the combined impact of compressions and magneto-dilations, as opposed to each separately, 
and there may be interesting self-organization principles that apply to combinations of $B$ and $\rho$. However, in this work, our focus on the weakly collisional model 
implies that magneto-immutability is subdominant to incompressibility (see next section). We thus consider the two effects separately, leaving
speculation about their interaction   in collisionless plasma turbulence to future work.

\subsection{Alfv\'enic turbulence with Braginskii viscosity}\label{sec:sub:alfvenic}
Although the arguments in the preceding paragraphs are quite general, 
we focus here on applying them to strong, Alfv\'enic turbulence \citep{Goldreich:1995hq} in the weakly collisional limit. 
We  define the turbulence amplitude $\dbturb\equiv\delta B_{\perp}/B_{0}\simc\delta u_{\perp}/\valf$ and the 
Alfv\'en frequency $\omA=k_{\parallel}\valf$ (where $k_{\parallel}^{-1}\simc l_{\parallel}$ is the field-parallel scale of a given fluctuation, and $k_{\perp}^{-1}$ is its perpendicular scale). 
We assume that $\beta\,\gg\nu_{c}/\omA\gg\beta^{1/2}\gg{1}$ (or, equivalently, $\beta^{-1/2}\ll k_{\parallel}\lambda_{\mathrm{mfp}}\ll1$), so that the ion-collision timescale $\nu_{c}^{-1}$ is longer than all other time scales, including those associated with $\mathcal{Q}(q_{\perp},q_{\parallel})$ \citep{Mikhailovskii:1971,Squire:2016ev2}. The result is a closure for $\Dp$ in which $\Dp$ is smaller than the variation in $p_{\perp}$ or $p_{\parallel}$ individually.
Equation~\eqref{eq:MHD.Dp}  becomes   
\begin{equation}
\Dp\approx\frac{p_{0}}{\nu_{c}}\,\bbgu,\label{eq:dp.brag}
\end{equation}
 where $\Dp\ll\!p_{\perp}\simeq\!p_{\parallel}\simeq p_{0}$ \citep{Braginskii:1965vl}. 
Because $\beta\gg1$, the flow is nearly incompressible and $p_{0}\!\simeq\const$~in \eqref{eq:dp.brag}. The pressure-anisotropy stress then takes the form of a field-aligned viscous stress $\visbrag\grad\bcdot[\bh\bh\,(\bbgu)]$, where 
$\visbrag\equiv\nu_{c}^{-1}p_{0}$ is the Braginskii viscosity. This model is thus
 often called ``Braginskii MHD.''
As discussed in \S\ref{sec:sub:interruption.description}, intuitively, we expect a strong modification of the turbulence for amplitudes above which shear Alfv\'en waves are interrupted and cannot propagate: $\dbturb\gtrsim\dbint\equiv\!2\beta^{-1/2}\sqrt{\nu_{c}/\omA}$. Note
that, because $\nu_{c}\gg \omA$, a weakly collisional plasma with fluctuations that satisfy $\dbturb\gtrsim\dbint$ necessarily also has $\beta\gg1$, justifying our use of an incompressible model in \S\ref{sec:sims} below. 

Because $\Dp\propto\bbgu$, the Braginskii viscous stress acts in the direction required to make the flow magneto-immutable. 
The fact that it irreversibly dissipates kinetic energy (unlike, for example, the pressure force $-\grad p$) is not important for our arguments here. 
A direct analogy for compressional motions is the bulk viscosity, which
has the form $-\mu_{\mathrm{bulk}}\grad(\grad\bcdot\bm{u})$ and damps compression and rarefaction of the flow. 
Interestingly, flows with large bulk viscosities (which are not commonly studied) are effectively incompressible
even when the Mach number based on the thermal pressure is large  \citep{Pan:2017gt}. 

By analogy with the Reynolds number -- which is the ratio of viscous to inertial time scales, \emph{viz.}, $\mathrm{Re}=\rho\delta u_{\perp}l_{\perp}/\mu_{\mathrm{iso}}\simc \rho\valf l_{\parallel}/\mu_{\mathrm{iso}}$ in MHD turbulence  (with isotropic dynamic viscosity $\mu_{\mathrm{iso}}$) -- we define the
Braginskii ``interruption number'' $\reyint$. $\reyint$ is the ratio of the timescale for the parallel viscous stress to act on an Alfv\'enically polarized motion,\footnote{Equivalently, this is the timescale for  $|\Dp|$ to change by ${\sim}B^{2}$ in an Alfv\'enic motion.} $t_{\mathrm{int}}\sim\omA^{-1}\dbint^{2}/\dbturb^{2}$ (see \citealt{Squire:2016ev2}), 
and the inertial timescale, $t_{\mathrm{inertial}}\sim (k_{\perp}\delta u_{\perp})^{-1} \sim (k_{\parallel}\valf)^{-1}$ (assuming critically balanced turbulence; \citealp{Goldreich:1995hq,Goldreich1997}), giving
\begin{equation}
\reyint\equiv \frac{t_{\mathrm{int}}}{t_{\mathrm{inertial}}}\approx\frac{\dbint^{2}}{\dbturb^{2}}\sim  \frac{\rho\valf l_{\parallel}}{\visbrag}\left(\frac{\delta B_{\perp}^{2}}{B_{0}^{2}}\right)^{-1}.\label{eq:reyint.definition}
\end{equation}
 The Braginskii stress will be dynamically important, i.e., comparable to the Maxwell and Reynolds stresses, $\bm{B}\bcdot\grad\bm{B}$ and $\bm{u}\bcdot\grad\bm{u}$,  for $\reyint\lesssim 1$, or  equivalently $\dbturb\gtrsim\dbint$.
As discussed above, when $\visbrag$ is so large  that $\reyint\lesssim1$, motions  become increasingly magneto-immutable, limiting $\Dp$ fluctuations to $\Dp\simc B^{2}$, in order to balance $\bm{B}\bcdot\grad\bm{B}$. Thus, when $\reyint\ll{1}$, keeping the amplitudes of $\bm{u}$ and $\bm{B}$ fluctuations approximately constant and changing $\visbrag$, we expect $(\bbgu)_{\mathrm{rms}}\propto\reyint$, or $\Dp_{\mathrm{rms}}\sim \mathrm\const$,  as opposed to the naive scaling, $(\bbgu)_{\mathrm{rms}}\sim\const$,  or $\Dp_{\mathrm{rms}}\propto\reyint^{-1}$, which holds  at $\reyint\gg1$ when pressure-anisotropy forces play no role. 
Note that in realistic plasmas, where microinstabilities can break the direct proportionality between $\bbgu$ and $\Dp$ (see \S\ref{sec:sub.microinstabilities} below), these 
scalings hold only in  regions that are not affected by microinstabilities.\footnote{For example, in our simulations reported below that use a mirror limiter, we measure $\Dp_{\mathrm{rms}}^{<0}\equiv\langle\Dp^{2}|_{\Dp<0}\rangle^{1/2}$ to exclude mirror-limited regions; see \S\ref{sec:sub:results}.}



%
%
%
\subsection{Microinstabilities}\label{sec:sub.microinstabilities}
Sufficiently non-Maxwellian distribution functions are unstable to 
kinetic plasma instabilities, complicating the  
 arguments above and breaking the correspondence
 between compression/rarefaction and magneto-dilation. 
In the high-$\beta$ regime, the most relevant microinstabilities are the firehose \citep{Rosenbluth:1956} and mirror \citep{Barnes1966,1969PhFl...12.2642H}, which are triggered when $\Dp\lesssim -B^{2}/4\pi$ and $\Dp\gtrsim B^{2}/8\pi$, respectively.
These instabilities act to deplete the amount of large-scale $\Dp$ in excess of the stability thresholds ($|\Dp|\lesssim B^2/4\pi$; \citealp{Schekochihin:2008en,Hellinger:2008hd,Kunz:2014kt,Melville:2015tt}), 
which they achieve over short time scales set by $\Omega_{i}$.
They may thus frustrate the plasma's attempts to become magneto-immutable by 
truncating the growth of $\Dp$ when it
becomes too large. 
There is no  analogue to this effect 
in (collisional) compressible hydrodynamic
flows, which are generally not strongly affected by kinetic instabilities because large variations in isotropic pressure can 
 occur even when $\nu_{c}^{-1}$ 
  is small compared to all other time scales (unlike $\Dp$, which is always negligibly small at sufficiently small $\nu_{c}^{-1}$).
Nonetheless, we argue, and show explicitly below (figure \ref{fig:pdfs}), that magneto-immutability remains an important
self-organizing principle, even if mirror and firehose perfectly limit $\Dp$ (i.e., $-B^{2}/4\pi\le\Dp\le B^{2}/8\pi$). 
The reason is that the two effects, microinstabilities and magneto-immutability, scale in 
identical ways: they are both important only once $\Dp\simc B^{2}$, implying 
that the limiting effect of microinstabilities does not dominate over magneto-immutability, or vice versa.

\section{Braginskii-MHD simulations}\label{sec:sims}

We now supplement the heuristic arguments proposed above by numerical simulations of Alfv\'enic turbulence. 
We use incompressible Braginskii MHD (equations \eqref{eq:MHD.u}--\eqref{eq:MHD.b} with $\Delta p$ given by \eqref{eq:dp.brag}) because it is the simplest model 
that captures the pressure-anisotropy effects of interest, allowing comparatively straightforward diagnosis of the key physics. 
The results of these simulations demonstrate three key points:
(i) that magneto-immutable turbulence with $\reyint\lesssim1$ ($\dbturb\gtrsim\dbint$) is possible and similar to 
standard critically balanced Alfv\'enic MHD turbulence (although some key differences do exist); (ii) that the pressure-anisotropy stress does indeed act to minimize $\bbgu$; and (iii) that the system approaches a well-defined nonzero turbulent state in the $\reyint\rightarrow0$ limit, 
similarly to the way in which hydrodynamic turbulence approaches incompressibility in the low-Mach-number limit.

\subsection{Numerics}\label{sec:sub:numerics}
Our simulations use the \textsc{Snoopy} code \citep{Lesur:2007bh},
which is based on a Fourier pseudo-spectral discretization in space. 
The pressure anisotropy $\Dp$ is calculated  from \eqref{eq:dp.brag}, with sub-cycling of the final term in \eqref{eq:MHD.u} eight times per global MHD timestep. The effect of microinstabilities is modeled
 by  limiting the value of $\Dp$ \citep{Sharma:2006dh}, \emph{viz.}, $\Dp=\min(\visbrag\bbgu,\,B^{2}/8\pi)$ (mirror) or 
 $\Dp=\max(\visbrag\bbgu,\,-B^{2}/4\pi)$ (firehose).  
Because the parallel firehose instability is captured by the Braginskii MHD model but the mirror instability is not, most simulations use only a mirror limiter. This choice also  
helps us to isolate  the  effects of magneto-immutability from those of the limiter, because $\Dp$  freely evolves in regions where  $\Dp<0$.
  However, we acknowledge that some crucial aspects of the 
   true kinetic firehose instability -- in particular, pitch-angle scattering of particles from ion-Larmor-scale fluctuations -- are not captured by Braginskii MHD. For this reason, we also run some turbulence simulations with both a mirror and a firehose limiter, which show similar qualitative behaviors to those with just a mirror limiter.
We use periodic boundary conditions in a three-dimensional box threaded by a uniform mean magnetic field $\bb{B}_0 = B_0\bb{\hat{x}}$. In all cases, $L_y/L_z=1$, whereas $L_x/L_z$ is varied depending upon the amplitude of the turbulent fluctuations. The latter are driven by forcing all modes of the velocity field up to ($|k_{x}|=2\times2\pi/L_{x}$, $|k_{y}|=2\times2\pi/L_{y}$, $|k_{z}|=2\times2\pi/L_{z}$)
 using an Orstein-Uhlenbeck process with correlation time $\simc\tau_{A}\equiv L_{x}/\valf$. The amplitude of the driving is chosen such that $\dbturb\equiv\delta B_{\perp}/B_{0}\simc \delta u_{\perp}/\valf\simc L_{y}/L_{x}=L_{z}/L_{x}$; i.e., we drive turbulence in critical balance, $k_{\parallel}\valf\simc k_{\perp}u_{\perp}$ \citep{Goldreich:1995hq}. 
 We present results for both trans-Alfv\'enic turbulence, with $L_{x}=L_{z}$ ($\dbturb\approx1$), and sub-Alfv\'enic turbulence in a box that is elongated along the mean-field direction, with $L_{x}=4L_{z}$ ($\dbturb\approx1/4$). We use fourth-order isotropic hyper-dissipation in $\bm{u}$ and $\bm{B}$ ($\mu_{\mathrm{iso,}4}\nabla^{4}\bm{u}$ and $\eta_{4}\nabla^{4}\bm{B}$), which 
 was chosen, after extensive testing with MHD simulations, because it gave the cleanest inertial range at a given resolution. Simulations are run until  $t=4\tau_{A}$ and results are averaged over the final $2\tau_{A}$. 
 
\begin{figure}
\begin{center}
\includegraphics[width=1\columnwidth]{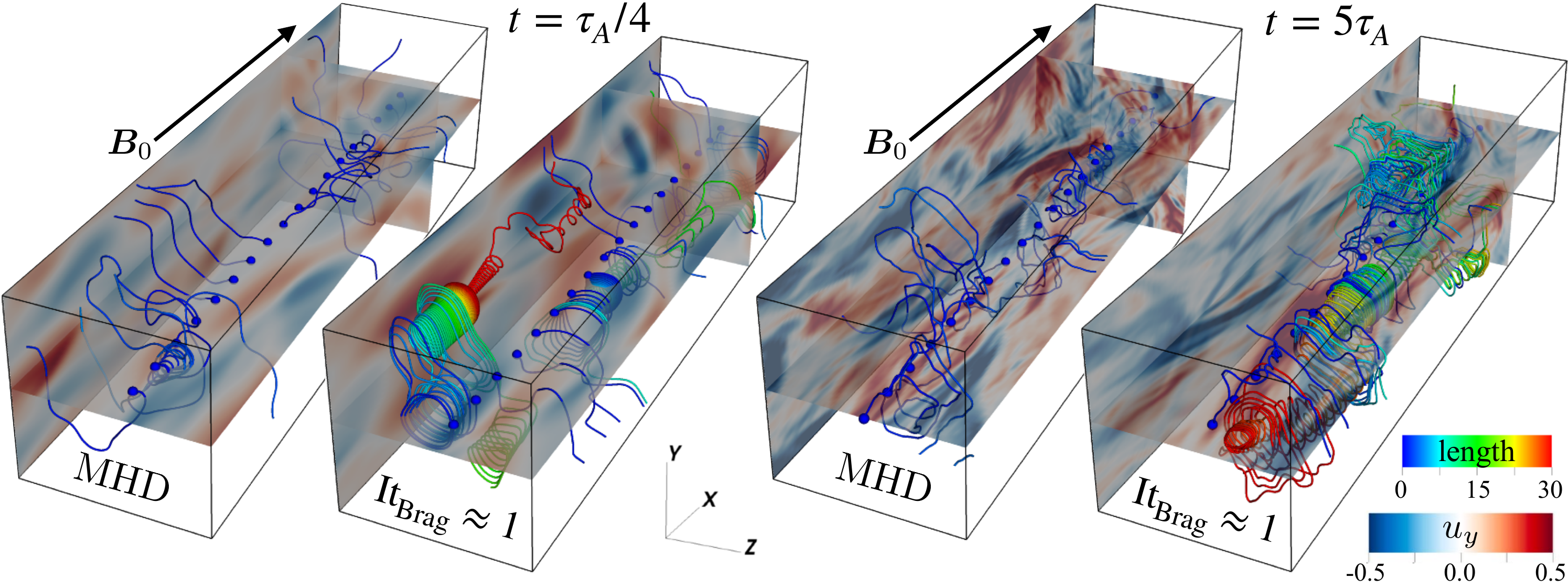}
\caption{The effect of pressure-anisotropy stress on the flow structure. In each panel, the color scale on each 
slice shows $u_{y}$ (perpendicular to $\bm{B}_{0}$), while the lines follow the streamlines of the incompressible flow (the color shows the length of a streamline from its origin, to more clearly show the different flow structures in each case). We compare MHD and  $\reyint=1$ flows, driven with identical forcing fields from zero initial conditions. 
The pair of panels on the left is for $t=\tau_{A}/4=0.25 L_{x}/\valf$, at which point the flow 
has not yet become fully turbulent, while the  panels on the right show  turbulent flow structures at $t=5\tau_{A}=5 L_{x}/\valf$. In both cases, the effect of magneto-immutability  is clearly seen in the flow lines, which become more tightly curled so that the flow has the direction required to avoid changes in $B$. This is a nonlinear analogue of a circularly polarized Alfv\'en wave.}
\label{fig:streamlines}
\end{center}
\end{figure}
 
We change the relative importance of the pressure-anisotropy stress
 by varying $\visbrag$ at constant forcing amplitude and constant $B_{0}$. As explained in \S\ref{sec:general.ramblings}, we expect pressure anisotropy
 to be important when $\reyint\lesssim1$, or equivalently  for $\visbrag\gtrsim\pi^{-1}\rho\valf L_{x}^{3}/L_{z}^{2}$ [see \eqref{eq:reyint.definition}]. Unfortunately, such a    large $\visbrag$ requires very short timesteps. Consequently, simulations at small $\reyint$ are vastly 
 more  expensive computationally than their MHD counterparts, and our highest resolutions are rather modest: $N_{x}=N_{y}=N_{z}=N_{x,y,z}=192$. Although other numerical methods may
 enable increased resolution in future work, great care must be taken: due to the large 
 values of $\visbrag$, very small errors in evaluating $\bbgu$ can spuriously damp legitimate motions. We chose the  pseudo-spectral  method after extensive tests
 of decaying turbulence with $\reyint>1$ but large $\visbrag$, using a variety of different numerical methods. In particular, unexpected problems arose in evaluating the Braginskii stress using finite-volume, operator-split methods.


\subsection{Results}\label{sec:sub:results}
To illustrate a magneto-immutable flow, in figure \ref{fig:streamlines} we compare the flow streamlines at early times using $\reyint\approx1$ Braginskii MHD with those obtained using standard MHD. Although the magnitude of 
the velocity in each case is  similar, the magneto-immutable flow has manifestly different structure: plasma is constrained to flow
along the direction that minimizes changes in $B$. The dynamics illustrated in figure \ref{fig:streamlines} may be thought of as a nonlinear generalization of a circularly polarized linear Alfv\'en wave, which does not change the strength of $B$.

\begin{figure*}
\begin{center}
\includegraphics[width=0.49\textwidth]{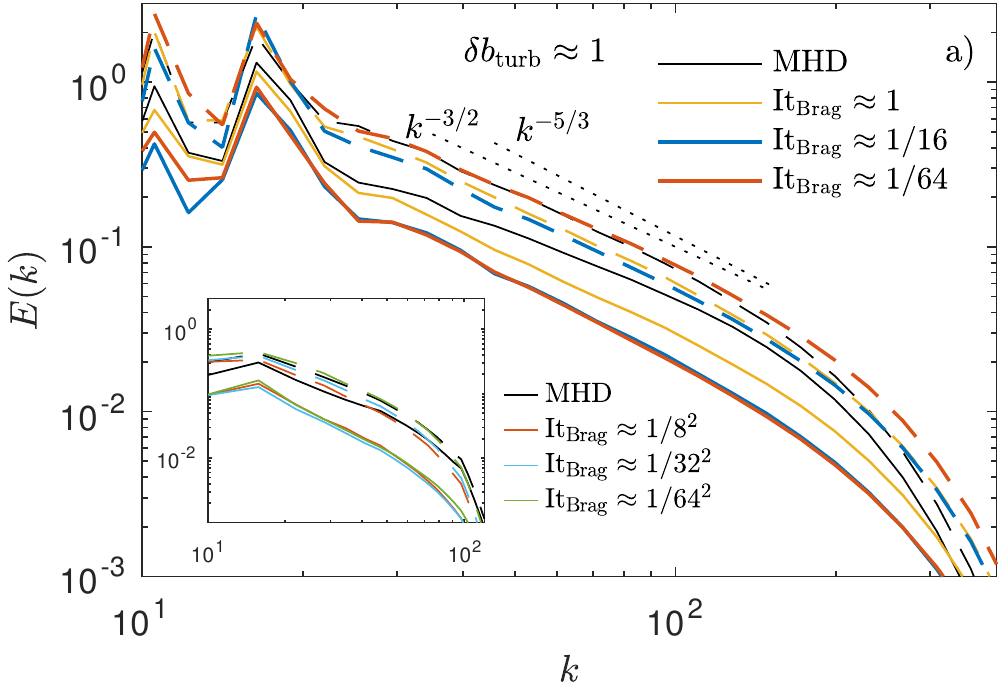}~~\includegraphics[width=0.49\textwidth]{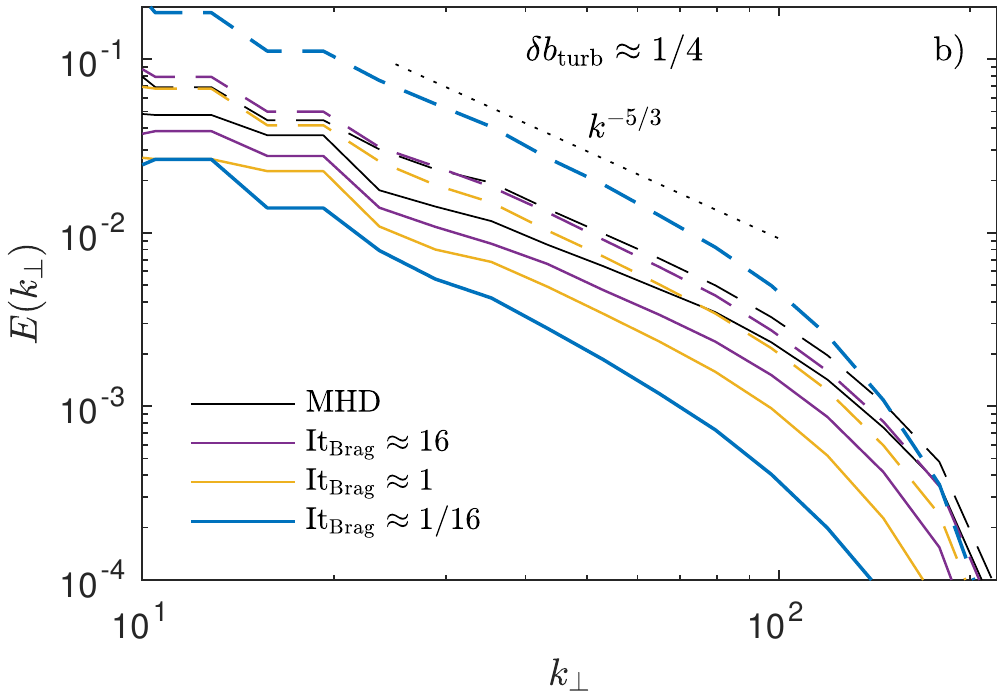}\\
\includegraphics[width=0.49\textwidth]{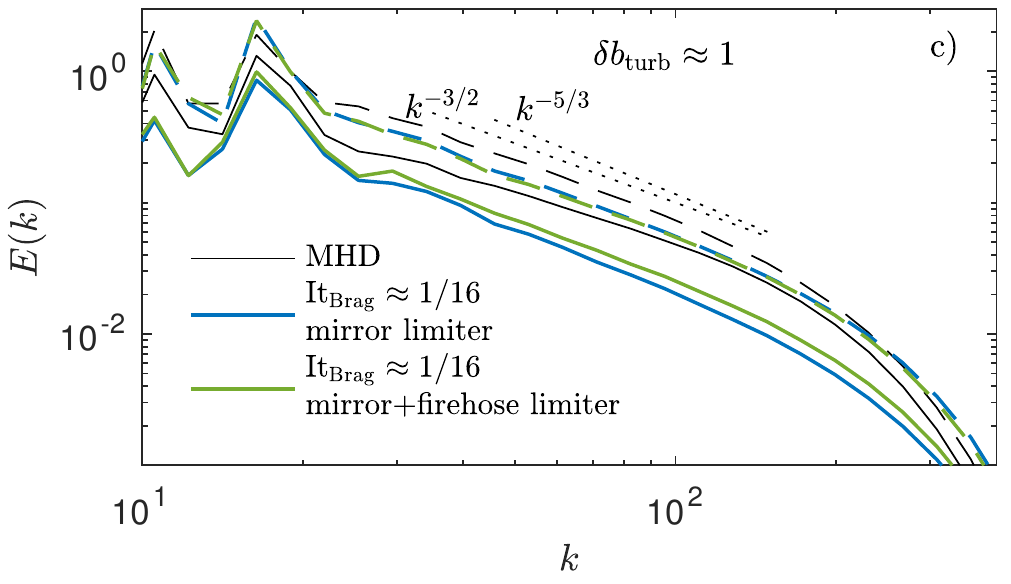}~~\includegraphics[width=0.49\textwidth]{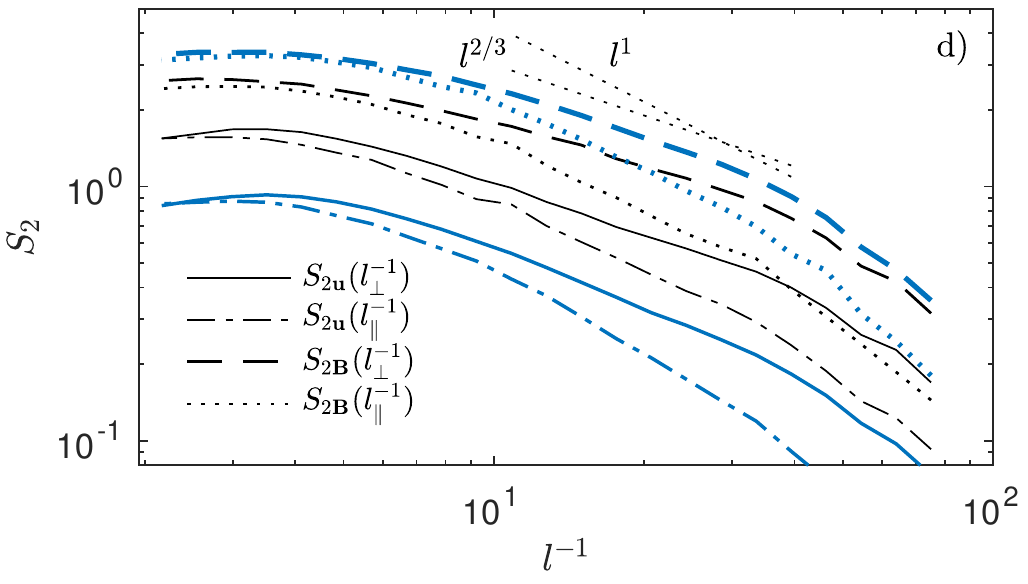}
\caption{Panel (a): kinetic-energy ($E_{K}$, solid lines) and magnetic-energy ($E_{M}$, dashed lines) spectra, for trans-Alfv\'enic turbulence simulations
with $\dbturb=\delta B_{\perp}/B_{0}\approx1$  ($L_{x}=1$) at resolution $N_{x,y,z}=192$, and a mirror limiter but no firehose limiter. As labeled, the different colors show simulations with different  $\reyint\approx\dbint^{2}/\dbturb^{2}$.  We expect the turbulence 
to be affected by magneto-immutability for $\reyint\lesssim1$. The inset shows 
spectra for low-resolution simulations ($N_{x,y,z}=48$) at even smaller $\reyint$.
Panel (b):
spectra for sub-Alfv\'enic turbulence simulations, $\dbturb=\delta B_{\perp}/B_{0}=1/4$ ($L_{x}=4$) and resolution $N_{x,y,z}=96$ (we bin energies in $k_{\perp}=(k_{y}^{2}+k_{z}^{2})^{1/2}$ in this case due to the elongated box). 
Panel (c): as in  (a) (trans-Alfv\'enic turbulence, $L_{x}=1$), but comparing the case with only a mirror limiter (blue) to that with both a mirror and firehose limiter (green) for  $\reyint\approx1/16$. The two are very similar, with a slightly smaller residual energy when the firehose limiter is used. 
Panel (d): anisotropic structure functions of the magnetic and kinetic energy ($S_{2\bm{B}} = \langle [\bm{B}(\bm{x}+\bm{l})-\bm{B}(\bm{x})]^{2}\rangle$ and $S_{2\bm{u}} = \langle [\bm{u}(\bm{x}+\bm{l})-\bm{u}(\bm{x})]^{2}\rangle$, respectively) for  trans-Alfv\'enic turbulence ($L_{x}=1$). Blue curves show  $\reyint\approx1/16$ Braginskii MHD turbulence while black curves show MHD (we plot $S_{2}$ versus $l^{-1}$ for  comparison with the other panels). The increments $\bm{l}$ are taken either perpendicular to the local scale-dependent 
magnetic field, $S_{2}({l}_{\perp}^{-1})$, or parallel to the field, $S_{2}({l}_{\parallel}^{-1})$, illustrating  increasing anisotropy at small scales, as in MHD \citep{Goldreich:1995hq,Goldreich1997}.}
\label{fig:spectra}
\end{center}
\end{figure*}

We now describe the key findings of our turbulence simulations (illustrated in Figs.~\ref{fig:spectra}--\ref{fig:ER.scaling}) and how these add to the discussion of \S\ref{sec:general.ramblings}.

\subsubsection{Turbulence is possible and Alfv\'enic in character}As discussed in \S\ref{sec:sub:interruption.description}, it is not obvious that turbulent motions can be supported at all when $\dbturb\gtrsim\dbint$ ($\reyint\lesssim1$), because isolated linearly polarized Alfv\'enic fluctuations cannot propagate (even with mirror and/or firehose limiters; \citealp{Squire:2016ev}). 
Our first result, illustrated in figure \ref{fig:spectra}, is that Braginskii MHD \emph{can} sustain 
turbulence when $\reyint<1$. Energy spectra are similar to those in MHD, but with increasing turbulent residual energy, $ E_{R}\equiv [\langle(\bm{B}-\bm{B}_{0})^{2}\rangle-\langle\bm{u}^{2}\rangle]/[\langle(\bm{B}-\bm{B}_{0})^{2}\rangle\langle\bm{u}^{2}\rangle]^{1/2}$, at low $\reyint$ (i.e., the system becomes more magnetically dominated, as occurs in an interrupted shear Alfv\'en wave). 
Spectral slopes are close to $k^{-5/3}$, or slightly
shallower (cf.~\citealt{Maron:2001cs,Boldyrev:2006ta,Perez:2012da,Beresnyak:2012ek}).
Comparing figures \ref{fig:spectra}({\it a}) and \ref{fig:spectra}({\it b}), we see that trans-Alfv\'enic and sub-Alfv\'enic turbulence are broadly similar at the same $\reyint$, \emph{viz.,}  $\dbint=1/4$ turbulence with $\dbturb\approx1/4$ is comparable to $\dbint=1$ turbulence with $\dbturb\approx1$ (although 
the  residual energy is larger  in the sub-Alfv\'enic case). We also see, in figure \ref{fig:spectra}({\it c}), that $\reyint<1$ turbulence with both 
mirror and firehose limiters  on $\Dp$ is relatively similar to that with just a mirror limiter, aside from the slightly smaller $ E_{R}$. 

We have
 run a variety of other common MHD-turbulence diagnostics on these simulation sets, including calculations of anisotropic structure functions of the kinetic and magnetic energy, which are shown in figure \ref{fig:spectra}({\it d}) for the trans-Alfv\'enic MHD and $\reyint=1/16$ simulations. These  are calculated using the method of \citet{2011MNRAS.415.3219C} and \citet{2015MNRAS.449L..77M}, by selecting for increments $\bm{l}$ that are either perpendicular ($\cos^{-1}(\hat{\bm{l}}\bcdot\hat{\bm{b}})>70^{\circ}$) or parallel ($\cos^{-1}(\hat{\bm{l}}\bcdot\hat{\bm{b}})<20^{\circ}$) to the local magnetic field around the chosen increment  $\bm{B}[(\bm{x}_{1}+\bm{x}_{2})/2]$, where $\bm{l}=\bm{x}_{2}-\bm{x}_{1}$. 
 We clearly see the signatures of scale-dependent anisotropy in both simulations,
 with the cascade following the  scalings $S_{2}\sim l_{\perp}^{2/3}$ and $S_{2}\sim l_{\parallel}^{1}$ usually expected
 for a critically balanced MHD cascade. Note that this calculation is carried out on the trans-Alfv\'enic simulations in a cubic 
 box with isotropic forcing,
 so the anisotropy measurement is not influenced by the assumption of critical balance in the outer-scale forcing.
We have also computed the alignment of $\bm{u}$ and $\bm{B}$ \citep[using the method of][]{2016MNRAS.459.2130M}, again finding  no striking differences compared to MHD turbulence (not shown). 

Overall, the biggest difference compared to MHD is the increase in $ E_{R}$. This appears to be related, in part, to $\langle \Dp\rangle$ being negative (thus changing the ratio of $\delta u_{\perp}$ to $\delta B_{\perp}$ in an Alfv\'en wave), as well as to the extra dissipation  in the momentum equation (but not the induction equation) due to Braginskii viscosity (see figure \ref{fig:ER.scaling}(a)). However, the behavior of $ E_{R}$, including why its relative increase is larger in sub-Alfv\'enic than trans-Alfv\'enic turbulence,  is not well understood by us at the present time. More generally, aside from these  differences in $ E_{R}$, 
it remains  unclear how $\reyint<1$ 
turbulence can be so \emph{similar} to MHD turbulence. The magnitude of the velocity fluctuations remains well above the interruption limit in all $\reyint<1$ simulations (and for $\reyint\ll1$, severely so), 
implying that isolated linearly polarized Alfv\'enic fluctuations would be unable to propagate for amplitudes similar to those  we find in our turbulence (see \S\ref{sec:sub:interruption.description} for further discussion).
Evidently, further study of other statistics and the structures in the flow and magnetic field is warranted (see, e.g., \citealt{Perez2009,Zhdankin2016}). However, given the limited resolution of our simulations, we leave this to future work.

\begin{figure}
\begin{center}
\includegraphics[width=0.6\columnwidth]{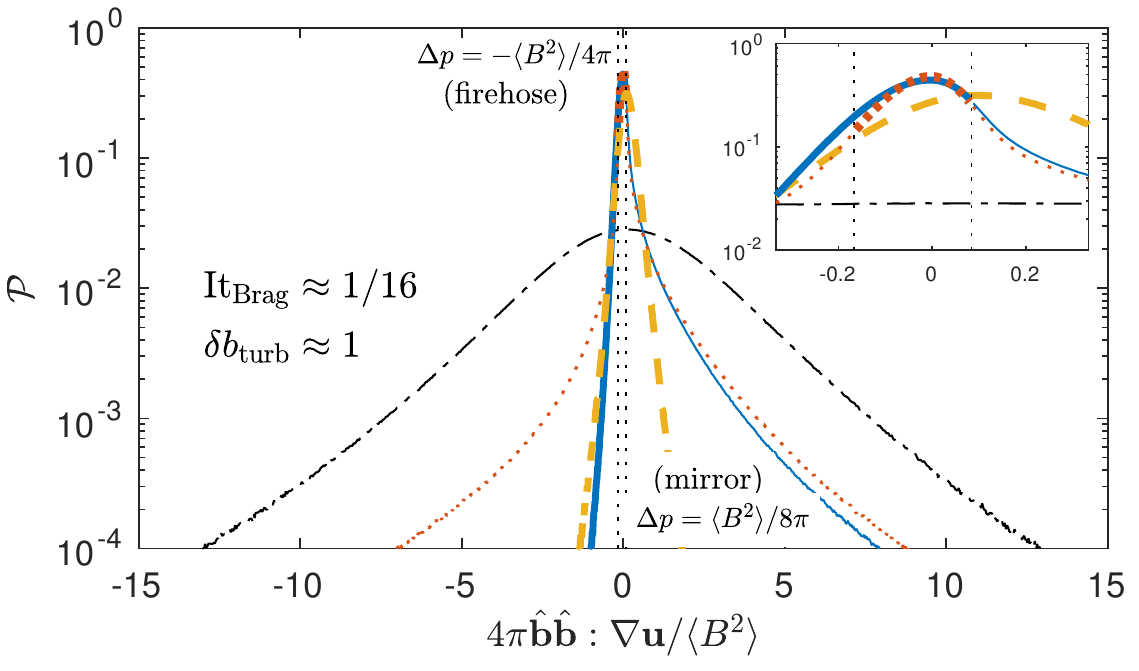}
\caption{PDF of $4\pi\,\bbgu/\langle B^{2}\rangle$ for $\dbturb\approx1$ ($L_{x}=1$) simulations (spectra shown in figure \ref{fig:spectra}). We 
compare MHD turbulence (black-dot-dashed line) to  $\reyint\approx1/16$  ($\dbint\approx1/4$, $\visbrag\approx2$) turbulence with a mirror limiter  (blue line), with both mirror and firehose limiters (red-dotted line), and with no limiters (yellow-dashed line).  The vertical dotted
lines denote the mirror and firehose limits for the $\reyint\approx1/16$ simulations.
 Regions with thicker lines (e.g., $\bbgu$ below the mirror limit for the blue line, or $\bbgu$ between the firehose and mirror limits for the red-dotted line) indicate where pressure-anisotropy 
forces are dynamically relevant (not limited).  The inset is a zoom into the central region.  This
figure shows that magneto-immutability forces significantly decrease the probability of 
 turbulence producing large changes in magnetic-field strength. Note that the change in 
$\langle B^{2}\rangle$ between these simulations is modest, and not the cause of the significant changes to the width of the PDF.
 %
 %
 %
 }
\label{fig:pdfs}
\end{center}
\end{figure}

The spectra and structure functions shown in figure \ref{fig:spectra} are specific to Braginskii MHD with microinstability limiter(s). Although an exhaustive survey is not the purpose of this work, it is helpful to briefly comment on their robustness.  Spectral slopes
and general features (e.g., scale-dependent anisotropy) are robust to changing 
the mirror-limit threshold, although, like the addition of a firehose limit (figure \ref{fig:spectra}({\it c})), these modifications result in  modest 
changes in the residual energy at 
a given $\reyint$. In the unphysical case without microinstability limiters -- i.e., 
when $\Dp$ is completely free to evolve -- the characteristics of the turbulence differ further, because
 $\langle\Dp\rangle$ is tied 
directly to the dissipation of $\bm{B}$, thus driving $\langle\Dp\rangle>0$ (see figure \ref{fig:pdfs}).\footnote{More precisely, if 
$\bm{B}$ had small-scale structure and its statistics were constant in time, then $\langle\bbgu\rangle$ would be positive (to see this, compute $\langle D\ln B/Dt \rangle=\langle \bbgu\rangle + \eta_{4}\langle \bm{B}\cdot\nabla^{4}\bm{B}/B^{2}\rangle$, and note that the final dissipation term is negative; see also \citealt{2016JPlPh..82f9001H}). Thus, for the system to be turbulent, $\langle \Dp\rangle$ -- which is related to 
$\bbgu$ through $\Dp=\visbrag\bbgu$ -- must increase indefinitely with decreasing $\reyint$. This is no longer true with a mirror and/or firehose limiter, which breaks the proportionality between $\Dp$ and $\bbgu$. Thus, as well as being unphysical, turbulence with no limiters is fundamentally different to that with limiters (although it does share some similar features; see figure \ref{fig:pdfs}).}
Finally, because $\dbint$ depends on $k_{\parallel}$ in the weakly collisional regime  (through $\omA$) but not in the collisionless regime [see \eqref{eq:interruption.limit}], these  spectra are likely specific to Braginskii MHD. Further simulations are required to explore spectra in collisionless high-$\beta$ plasmas.

\subsubsection{Pressure-anisotropic forces reduce $\bbgu$}The key conjecture in \S\ref{sec:general.ramblings}, which we justified only heuristically, is that  pressure-anisotropy stresses inhibit motions with large magneto-dilations ($\bbgu$). That this is indeed the case is shown in figure \ref{fig:pdfs}, where we compare the probability density function (PDF) of $\bbgu$ in 
MHD turbulence and in Braginskii turbulence at $\reyint\approx1/16$ using both limiters, only a mirror limiter, or no limiters. 
We see that pressure-anisotropy forces are remarkably effective at preventing $|\bbgu|$ from becoming too large,  significantly 
reducing the range of $|\bbgu|$ produced by the turbulent motions.
Microinstability limiters -- which affect regions with $\Dp>B^{2}/8\pi$ and/or  with $\Dp<-B^{2}/4\pi$ -- increase the probabilities of larger $|\bbgu|$ because they sever the adiabatic tie between $\bbgu$ and the pressure anisotropy.
However, we see that, even in limiter-affected regions, large $|\bbgu|$ events are much less probable. Indeed, while 
${\simeq}54\%$ of the volume lies within the stable region $-B^{2}/4\pi<\Dp<B^{2}/8\pi$ in the mirror-firehose limited turbulence (red-dotted line), only ${\simeq}3\%$ of the equivalent MHD turbulence (black-dot-dashed line) does. This shows that microinstabilities do not eliminate the plasma's tendency towards magneto-immutability, even if they instantaneously constrain $\Dp$ to lie within the stable range of values.

\begin{figure*}
\begin{center}
\includegraphics[width=0.495\textwidth]{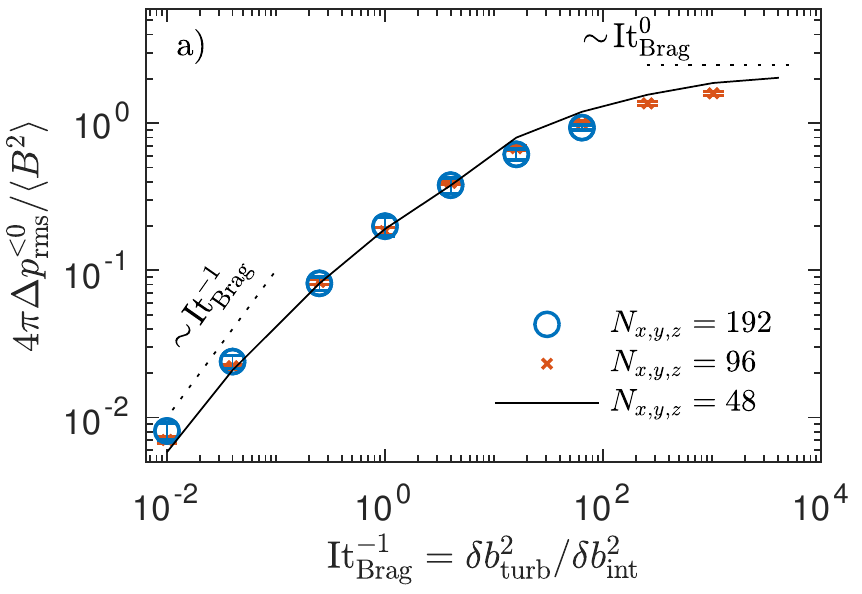}~\includegraphics[width=0.495\textwidth]{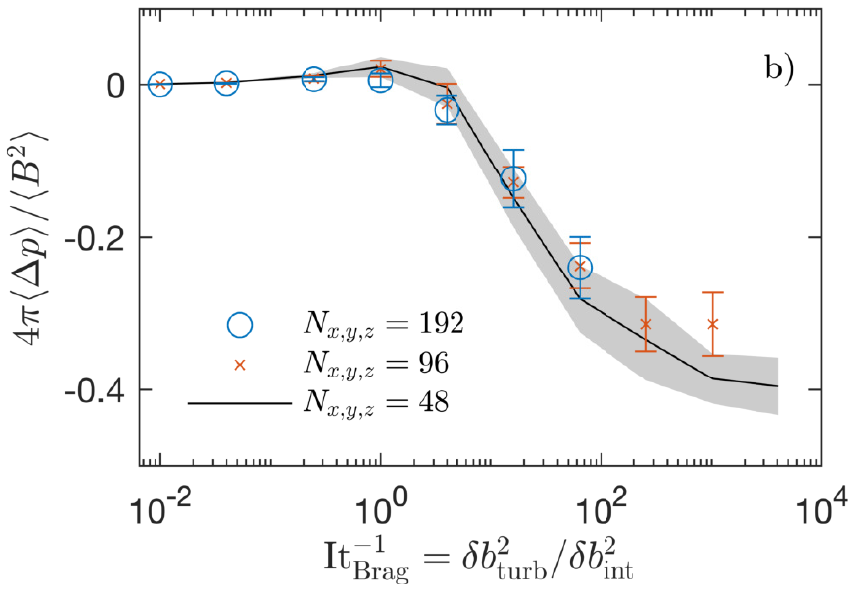}
\caption{Scaling  of $\Dp$  statistics with $\reyint$ in the $\dbturb\equiv\delta B_{\perp}/B_{0}\approx1$ ($L_{x}=1$) simulations with a mirror, but no firehose, limiter.
We compare simulation sets with varying resolution in order to explore the $\reyint\ll1$ regime of magneto-immutable turbulence.
(a) Width of the $\Dp$ distribution, calculated for $\Dp<0$, where the pressure anisotropy is not artificially limited ($\Dp_{\mathrm{rms}}^{<0}\equiv\langle\Dp^{2}|_{\Dp<0}\rangle^{1/2}$; see figure \ref{fig:pdfs}, thick blue line). The convergence of $4\pi\Dp_{\mathrm{rms}}^{<0}/\langle B^{2}\rangle$ to approximately $2$ at $\reyint\ll1$ shows that the 
flow becomes increasingly magneto-immutable with decreasing $\reyint$.
(b) Mean pressure anisotropy in each simulation, which also appears to converge to an asymptotic value $4\pi\langle \Dp\rangle\approx -0.4\langle B^{2}\rangle$ at $\reyint\ll1$.
Error bars  in each panel ishow the temporal dispersion of the plotted quantities.}
\label{fig:db.scaling}
\end{center}
\end{figure*}

\begin{figure}
\begin{center}
\includegraphics[width=0.495\textwidth]{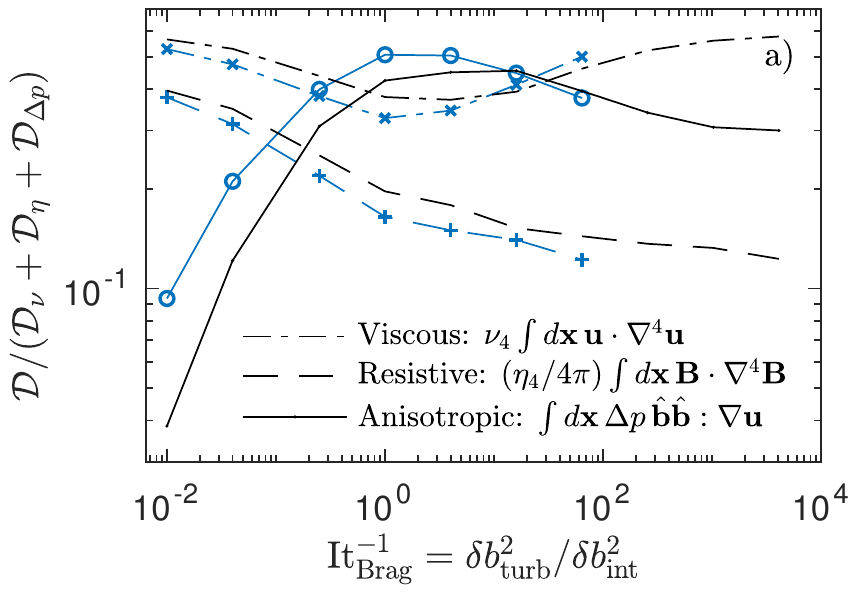}~\includegraphics[width=0.495\textwidth]{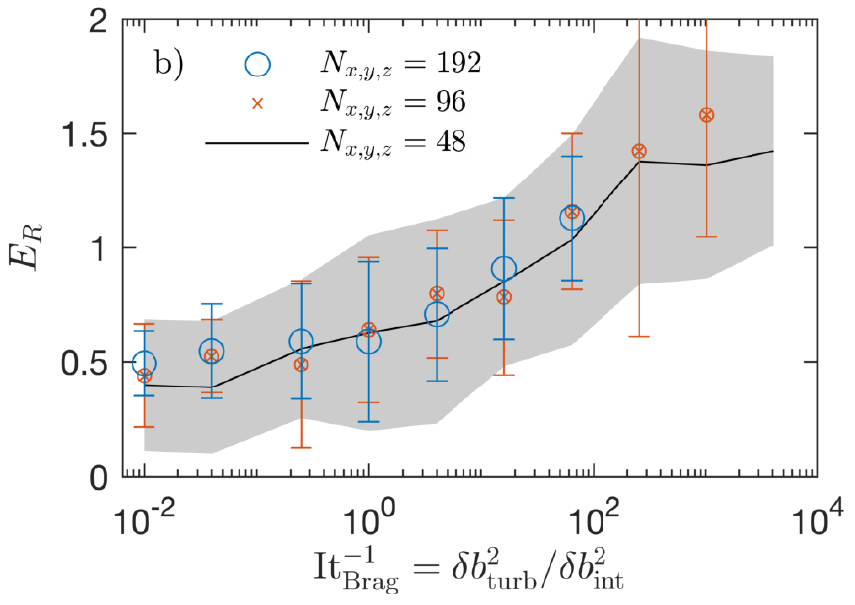}
\caption{Scaling properties of  turbulence statistics in the trans-Alfv\'enic ($\dbturb\equiv\delta B_{\perp}/B_{0}\approx1$) mirror-limited simulations, using the same conventions as figure~\ref{fig:db.scaling}. (a) Dissipation $\mathcal{D}$ (energy lost per $\tau_{A}$) due to $\Dp$ ($\mathcal{D}_{\Dp}\equiv\int\,d\bm{x}\,\Dp\,\bbgu$; solid lines), and due to hyper-viscosity 
($\mathcal{D}_{\nu}\equiv\nu_{4}\int\,d\bm{x}\,\bm{u}\bcdot\nabla^{4}\bm{u}$; dashed lines) and hyper-resistivity  ($\mathcal{D}_{\eta}\equiv(\eta_{4}/4\pi)\int\,d\bm{x}\,\bm{B}\bcdot\nabla^{4}\bm{B}$; dot-dashed lines), with values normalized to the total dissipation rate $\mathcal{D}_{\nu}+\mathcal{D}_{\eta}+\mathcal{D}_{\Dp}$. 
Blue lines with symbols show $N_{x,y,z}=192$ simulations, black lines show $N_{x,y,z}=48$ simulations. 
The anisotropic diffusion remains approximately constant for $\reyint\lesssim1$, 
despite the increasing $\visbrag$.
(b) Turbulent residual 
energy $ E_{R} $.
We see tentative evidence for the approach to an asymptotic value $ E_{R}\approx 1.5$ as  $\reyint\rightarrow0$, again suggesting that the turbulence has a well-defined magneto-immutable state for  
$\reyint \ll 1$.  }
\label{fig:ER.scaling}
\end{center}
\end{figure}

\subsubsection{The limit $\reyint{\rightarrow0}$ is well defined}An important assumption used in some arguments of \S\ref{sec:general.ramblings} is 
that an incompressible flow is \emph{able} to self-organize to minimize $\bbgu$, \emph{viz.}, that the system can approach 
a well-defined asymptotic state with non-zero $\bm{u}$ and $\bm{B}$ as  $\reyint\rightarrow0$. 
Figures~\ref{fig:db.scaling} and \ref{fig:ER.scaling} provide numerical evidence that this is the case. 
In particular, we see that key statistical properties of the turbulence appear to reach an asymptotic regime as $\reyint$ decreases. 
 Figure \ref{fig:db.scaling}({\it a}) shows that the width of the $\Dp$ distribution 
changes from scaling as $(\Dp)_{\mathrm{rms}}\sim\reyint^{-1}$ for $\reyint\gg1$, to $(\Dp)_{\mathrm{rms}}\sim\const$ when $\reyint\ll1$.
As discussed below [see \eqref{eq:reyint.definition}], 
this scaling demonstrates that  pressure-anisotropy forces  decrease $\bbgu$ so that the 
pressure-anisotropy stress is always comparable to $\bm{B}\bcdot\grad\bm{B}$, even as $\visbrag$ increases. The turbulence thus becomes more and more magneto-immutable. 
We also show, in figure \ref{fig:db.scaling}({\it b}), the mean pressure anisotropy $4\pi\langle \Dp\rangle/\langle B^{2}\rangle$ as 
a function of $\reyint$. This  appears to approach $\langle \Dp\rangle\approx -0.4 \langle B^{2}\rangle/4\pi$
at $\reyint\ll1$.
Finally, in figure \ref{fig:ER.scaling}({\it a}), we compare the turbulent dissipation due to Braginskii viscosity, $\mathcal{D}_{\Dp}\equiv\int\,d\bm{x}\,\Dp\,\bbgu$, with that due to hyper-viscosity 
and hyper-resistivity.  Because 
$\bbgu$ is unaffected by magneto-immutability in mirror-limiter regions, while $(\Dp)_{\mathrm{rms}}$ remains approximately constant with $\reyint$, 
the fraction of energy dissipated by Braginskii viscosity remains approximately constant for $\reyint\lesssim1$.


 Finally, the existence of this asymptotic regime in the statistics of $\Dp$ as $\reyint\rightarrow0$ suggests that 
the system can reach a well-defined magneto-immutable turbulent state, where turbulence properties -- e.g., 
velocity and field statistics -- do not depend on $\reyint$. This is possible because the typical size of the Braginskii 
viscous stress in the momentum equation, $\grad\bcdot(\bh\bh\,\Dp)$,
can become independent of $\visbrag$. 
Similar ideas are widely applied to compressible hydrodynamic turbulence,  
where the properties of the velocity field become effectively independent of Mach number $\mathcal{M}$ for $\mathcal{M}\ll1$. 
We give tentative evidence that our simulations approach this asymptotic 
magneto-immutable turbulence in figure \ref{fig:ER.scaling}({\it b}),
which shows that the 
turbulent residual energy $ E_{R}$ appears to approach a constant value for $\reyint^{-1}\gtrsim300$.
However,  we caution that the details of this asymptotic state -- e.g., the value of $ E_{R}$ as $\reyint\rightarrow0$ -- depend on the limiters used and the Braginskii-MHD model. Furthermore, reaching this asymptotic state 
is  computationally very challenging due to the 
 enormous $\visbrag$, and our lowest $\reyint$ simulations may be suspect due to their very low resolutions ($N_{x,y,z}=48$). The study of detailed flow and field  structures  and/or  statistics (e.g., scale-dependent anisotropy) at such a low resolution is of questionable utility,  so 
  it remains an open question how the properties of the turbulence at asymptotically low $\reyint$ 
  differ from those at moderate $\reyint$ or in MHD
  (although it is worth noting that energy spectra at $\reyint\approx1/64^{2}$ are similar to those at lower $\reyint$; see the inset of figure \ref{fig:spectra}({\it a})).
There is also clearly much 
further work  needed in order 
to understand $\reyint\rightarrow0$ turbulence in less collisional plasmas where $\nu_{c}\lesssim\beta^{1/2}\omA$ and the Braginskii 
MHD model does not apply.

\section{Conclusions}

We propose that weakly collisional and collisionless plasma turbulence 
is often ``magneto-immutable'' -- that is, it self-organizes to resist changes to  $|\bm{B}|$ by minimizing $|\bbgu|$.
This occurs due to the pressure-anisotropy stress $\grad\bcdot(\hat{\bm{b}}\hat{\bm{b}}\Dp)$, somewhat analogously to the way in which bulk pressure
forces (and bulk viscosity) render fluids incompressible.  In Alfv\'enic turbulence, our focus here, the effect is relevant for all 
scales above the plasma's kinetic microscales, and for fluctuation
 amplitudes around and above the ``interruption limit''  \eqref{eq:interruption.limit}  \citep{Squire:2016ev}. 
 By analogy with the Reynolds number, we 
  define the turbulent ``interruption number'' $\reyint$, which is the ratio of the ``pressure-anisotropy timescale'' (the timescale required 
  to generate $|\Dp|\simc B^{2}$) to the inertial timescale of the turbulence.
Turbulence becomes magneto-immutable for $\reyint\lesssim 1$, which, for trans-Alfv\'enic fluctuations ($\delta B_{\perp}\simc B$), occurs when $\beta\gtrsim\nu_{c}/\omega_{A}$ in a weakly collisional plasma, or when $\beta\gtrsim1$ in a collisionless plasma.  
While kinetic  microinstabilities frustrate the plasma's attempts to become magneto-immutable  by breaking the adiabatic link between $\bbgu$ and $\Dp$, they cannot eliminate the effect, even if they instantaneously constrain $\Dp$ to lie within the region of stability ($|\Dp|/p_{0}\lesssim \beta^{-1}$).

We confirm these ideas using driven magnetized-turbulence simulations in
the weakly collisional Braginskii MHD model, which contains 
the key  physics without truly kinetic complications.
The resulting magneto-immutable turbulence strongly resembles Alfv\'enic MHD  turbulence, 
displaying similar 
energy spectra and  scale-dependent anisotropy, although it exhibits a somewhat larger residual energy. 
This similarity  is particularly surprising given that isolated linearly polarized shear Alfv\'en waves -- generally 
considered to be the building blocks of MHD turbulence -- would be interrupted 
and unable to propagate for  fluctuation amplitudes similar to those seen in the turbulence. 
To get around this, it appears that the turbulent  flow self organizes into a nonlinear analogue of circular polarization, with tightly curled flow structures that avoid changing $B$ (see figure \ref{fig:streamlines}).
Examination of the probability density function of $\bbgu$ (related to $\Dp$ through $\Dp=\visbrag\bbgu$ in 
 Braginskii MHD) shows that the turbulence  strongly reduces the 
probability of fluctuations that generate high $\bbgu$ compared to MHD, without significantly reducing amplitude of the $\bm{u}$ and $\bm{B}$ fluctuations themselves. This effect is analogous to  low-Mach-number hydrodynamic fluctuations self-organizing to reduce 
the probability of high $\grad\bcdot\bm{u}$. In 
the limit of very high $\beta$ ($\reyint\rightarrow0$ or $\visbrag\rightarrow\infty$), we see tentative evidence that the turbulence 
approaches a well-defined magneto-immutable state, 
where the statistics of $\bm{u}$ and $\bm{B}$ no longer depend on the Braginskii viscosity (i.e., $\reyint$). Again, this
is analogous to how the statistics of $\bm{u}$ become independent of Mach number as subsonic turbulence 
becomes incompressible. 


 
A promising application of the ideas discussed throughout this work 
would be to MHD-scale  turbulence
in the collisionless solar wind, although the characteristics of magneto-immutability in the collisionless regime are admittedly still to be investigated at the present time. While many studies have found that solar-wind turbulence is well described by MHD models 
\citep{Matthaeus20140154,2016JPlPh..82f5302C}, we 
predict a key difference: that the distribution of $\bbgu$ should be much narrower than 
what would be driven by unconstrained (non-magneto-immutable) fluctuations of similar amplitude (see figure \ref{fig:pdfs}).
Intriguing evidence for this can be found in observations that 
show   $\bm{B}$ fluctuations preferentially trace out the surface of a sphere, keeping $|\bm{B}|$ approximately constant (see, e.g., figure~4 of \citealt{2001P&SS...49.1201B}, as well as \citealp{Lichtenstein1980,1993JGR....98.1257T,Tsurutani1994,Riley1996}). A magneto-immutability-based 
explanation for this behavior differs somewhat from the recent work of \citet{Tenerani2018}, who argue that  constant-$B$ fluctuations arise directly from the parallel firehose instability.  It is, however, consistent with the work of \citet{Vasquez1998}, who 
saw constant-$B$ states emerging  in (hybrid) kinetic  simulations. Further work on collisionless plasmas, as well as some understanding of magneto-immutability in an imbalanced cascade, 
is necessary 
before making detailed comparisons to solar-wind data.

On the theoretical side, a thought-provoking (if esoteric) question, is whether it is possible to  formulate directly and solve the equations for a truly magneto-immutable fluid, just as the incompressible fluid equations constitute a valuable model for subsonic fluid dynamics.
There remain many open questions related to 
the structure of magneto-immutable turbulence -- for instance, how it is {able} to remain so similar to Alfv\'enic MHD turbulence -- which will require higher-resolution simulations to address in detail. 
It is also important 
to move beyond the incompressible, high-collisionality Braginskii MHD model used here, exploring the influence of
heat fluxes on pressure-anisotropy stresses \citep{Mikhailovskii:1971}, how magneto-immutability effects interact
with density fluctuations (i.e., compressibility), the physics of magneto-immutability in the collisionless regime, and the role of realistic microinstability evolution \citep[e.g.,][]{Kunz:2014kt,Melville:2015tt}. 
These questions can be tackled in future work using Landau-fluid models \citep{Snyder:1997fs,SantosLima:2014cn,2015JPlPh..81a3203S} and/or MHD-scale kinetic simulations.

\acknowledgments
We thank S.~Cowley, P.~Kempski, R.~Meyrand, and  M.~Strumik for enlightening discussions.  Support for J.S. was provided by the  Marsden Fund grant UOO1727  managed through the Royal Society Te Ap\=arangi, and by the Gordon and Betty Moore Foundation through Grant GBMF5076 to Lars Bildsten, Eliot Quataert and E. Sterl Phinney. 
The work of A.A.S was supported in part by grants from UK STFC (ST/N000919/1) and EPSRC (EP/M022331/1).
E.Q.~was supported by Simons Investigator awards from the Simons Foundation and  NSF grants AST 13-33612 and AST 17-15054. 
M.W.K.~was supported in part by NASA grant NNX17AK63G, US DOE Contract DE-AC02-09-CH11466, and an Alfred P.~Sloan Research Fellowship. 
This work used the Extreme Science and Engineering Discovery Environment (XSEDE), which is supported by National Science Foundation Grant No.~ACI-1548562. Computations were carried out on the Comet system at the San Diego Supercomputing Center, through allocation TG-AST160068. Some of the numerical calculations presented in this work were done on Caltech's  Wheeler cluster. 

\bibliographystyle{jpp}
\bibliography{bib_extrapapers,fullbib}

\begin{thebibliography}{48}
\expandafter\ifx\csname natexlab\endcsname\relax\def\natexlab#1{#1}\fi

\bibitem[Bale {\em et~al.\/}(2009)Bale, Kasper, Howes, Quataert, Salem \&
  Sundkvist]{Bale:2009de}
{\sc Bale, S.~D., Kasper, J.~C., Howes, G.~G., Quataert, E., Salem, C. \&
  Sundkvist, D.} 2009 Magnetic fluctuation power near proton temperature
  anisotropy instability thresholds in the solar wind. {\em \prl\/} {\bf 103},
  211101.

\bibitem[{Barnes}(1966)]{Barnes1966}
{\sc {Barnes}, A.} 1966 Collisionless damping of hydromagnetic waves. {\em
  \pfluid\/} {\bf 9}, 1483.

\bibitem[{Barnes} \& {Hollweg}(1974)]{Barnes1974}
{\sc {Barnes}, A. \& {Hollweg}, J.~V.} 1974 {Large-amplitude hydromagnetic
  waves}. {\em \jgr\/} {\bf 79}, 2302.

\bibitem[Beresnyak(2012)]{Beresnyak:2012ek}
{\sc Beresnyak, A.} 2012 Basic properties of magnetohydrodynamic turbulence in
  the inertial range. {\em \mnras\/} {\bf 422}, 3495.

\bibitem[Boldyrev(2006)]{Boldyrev:2006ta}
{\sc Boldyrev, S.} 2006 Spectrum of magnetohydrodynamic turbulence. {\em
  \prl\/} {\bf 96}, 115002.

\bibitem[{Borovsky}(2008)]{2008JGRA..113.8110B}
{\sc {Borovsky}, J.~E.} 2008 {Flux tube texture of the solar wind: Strands of
  the magnetic carpet at 1 AU?} {\em \jgrsp\/} {\bf 113}, A08110.

\bibitem[Braginskii(1965)]{Braginskii:1965vl}
{\sc Braginskii, S.~I.} 1965 Transport processes in a plasma. {\em Rev. Plasma
  Phys.\/} {\bf 1}, 205.

\bibitem[{Bruno} {\em et~al.\/}(2001){Bruno}, {Carbone}, {Veltri},
  {Pietropaolo} \& {Bavassano}]{2001P&SS...49.1201B}
{\sc {Bruno}, R., {Carbone}, V., {Veltri}, P., {Pietropaolo}, E. \&
  {Bavassano}, B.} 2001 {Identifying intermittency events in the solar wind}.
  {\em \planss\/} {\bf 49}, 1201--1210.

\bibitem[{Chen}(2016)]{2016JPlPh..82f5302C}
{\sc {Chen}, C.~H.~K.} 2016 {Recent progress in astrophysical plasma turbulence
  from solar wind observations}. {\em \jplp\/} {\bf 82}, 535820602.

\bibitem[{Chen} {\em et~al.\/}(2011){Chen}, {Mallet}, {Yousef}, {Schekochihin}
  \& {Horbury}]{2011MNRAS.415.3219C}
{\sc {Chen}, C.~H.~K., {Mallet}, A., {Yousef}, T.~A., {Schekochihin}, A.~A. \&
  {Horbury}, T.~S.} 2011 {Anisotropy of Alfv{\'e}nic turbulence in the solar
  wind and numerical simulations}. {\em \mnras\/} {\bf 415}, 3219.

\bibitem[Chew {\em et~al.\/}(1956)Chew, Goldberger \& Low]{CGL:1956}
{\sc Chew, C.~F., Goldberger, M.~L. \& Low, F.~E.} 1956 {The Boltzmann equation
  and the one-fluid hydromagnetic equations in the absence of particle
  collisions}. {\em Proc. R. Soc. London A\/} {\bf 236}, 112.

\bibitem[Goldreich \& Sridhar(1995)]{Goldreich:1995hq}
{\sc Goldreich, P. \& Sridhar, S.} 1995 Toward a theory of interstellar
  turbulence. {S}trong {A}lfv{\'e}nic turbulence. {\em \apj\/} {\bf 438}, 763.

\bibitem[{Goldreich} \& {Sridhar}(1997)]{Goldreich1997}
{\sc {Goldreich}, P. \& {Sridhar}, S.} 1997 Magnetohydrodynamic turbulence
  revisited. {\em \apj\/} {\bf 485}, 680.

\bibitem[{Hasegawa}(1969)]{1969PhFl...12.2642H}
{\sc {Hasegawa}, A.} 1969 {Drift mirror instability of the magnetosphere}. {\em
  \pfluid\/} {\bf 12}, 2642.

\bibitem[{Helander} {\em et~al.\/}(2016){Helander}, {Strumik} \&
  {Schekochihin}]{2016JPlPh..82f9001H}
{\sc {Helander}, P., {Strumik}, M. \& {Schekochihin}, A.~A.} 2016 {Constraints
  on dynamo action in plasmas}. {\em \jplp\/} {\bf 82}, 905820601.

\bibitem[{Hellinger} \& {Tr{\'a}vn{\'{\i}}{\v c}ek}(2008)]{Hellinger:2008hd}
{\sc {Hellinger}, P. \& {Tr{\'a}vn{\'{\i}}{\v c}ek}, P.~M.} 2008 {Oblique
  proton fire hose instability in the expanding solar wind: Hybrid
  simulations}. {\em \jgrsp\/} {\bf 113}, A10109.

\bibitem[Kasper {\em et~al.\/}(2002)Kasper, Lazarus \&
  Gary]{2002GeoRL..29.1839K}
{\sc Kasper, J.~C., Lazarus, A.~J. \& Gary, S.~P.} 2002 {Wind/SWE observations
  of firehose constraint on solar wind proton temperature anisotropy}. {\em
  \grl\/} {\bf 29}~(1), 1839.

\bibitem[Kulsrud(1983)]{Kulsrud:1980tm}
{\sc Kulsrud, R.~M.} 1983 {MHD description of plasma}. In {\em {Handbook of
  Plasma Physics}\/} (ed. R~N Sagdeev \& M~N Rosenbluth). Princeton University.

\bibitem[Kunz {\em et~al.\/}(2014)Kunz, Schekochihin \& Stone]{Kunz:2014kt}
{\sc Kunz, M.~W., Schekochihin, A.~A. \& Stone, J.~M.} 2014 Firehose and mirror
  instabilities in a collisionless shearing plasma. {\em \prl\/} {\bf 112},
  205003.

\bibitem[Lesur \& Longaretti(2007)]{Lesur:2007bh}
{\sc Lesur, G. \& Longaretti, P.~Y.} 2007 Impact of dimensionless numbers on
  the efficiency of magnetorotational instability induced turbulent transport.
  {\em \mnras\/} {\bf 378}, 1471.

\bibitem[{Lichtenstein} \& {Sonett}(1980)]{Lichtenstein1980}
{\sc {Lichtenstein}, B.~R. \& {Sonett}, C.~P.} 1980 {Dynamic magnetic structure
  of large amplitude Alfv{\'e}nic variations in the solar wind}. {\em \grl\/}
  {\bf 7}, 189.

\bibitem[{Mallet} {\em et~al.\/}(2015){Mallet}, {Schekochihin} \&
  {Chandran}]{2015MNRAS.449L..77M}
{\sc {Mallet}, A., {Schekochihin}, A.~A. \& {Chandran}, B.~D.~G.} 2015 {Refined
  critical balance in strong Alfv{\'e}nic turbulence}. {\em \mnras\/} {\bf
  449}, L77--L81.

\bibitem[{Mallet} {\em et~al.\/}(2016){Mallet}, {Schekochihin}, {Chandran},
  {Chen}, {Horbury}, {Wicks} \& {Greenan}]{2016MNRAS.459.2130M}
{\sc {Mallet}, A., {Schekochihin}, A.~A., {Chandran}, B.~D.~G., {Chen},
  C.~H.~K., {Horbury}, T.~S., {Wicks}, R.~T. \& {Greenan}, C.~C.} 2016
  {Measures of three-dimensional anisotropy and intermittency in strong
  Alfv{\'e}nic turbulence}. {\em \mnras\/} {\bf 459}, 2130--2139.

\bibitem[Maron \& Goldreich(2001)]{Maron:2001cs}
{\sc Maron, J. \& Goldreich, P.} 2001 Simulations of incompressible
  magnetohydrodynamic turbulence. {\em \apj\/} {\bf 554}, 1175.

\bibitem[Matthaeus {\em et~al.\/}(2015)Matthaeus, Wan, Servidio, Greco, Osman,
  Oughton \& Dmitruk]{Matthaeus20140154}
{\sc Matthaeus, W.~H., Wan, M., Servidio, S., Greco, A., Osman, K.~T., Oughton,
  S. \& Dmitruk, P.} 2015 Intermittency, nonlinear dynamics and dissipation in
  the solar wind and astrophysical plasmas. {\em \prsa\/} {\bf 373}, 20140154.

\bibitem[Melville {\em et~al.\/}(2016)Melville, Schekochihin \&
  Kunz]{Melville:2015tt}
{\sc Melville, S., Schekochihin, A.~A. \& Kunz, M.~W.} 2016
  Pressure-anisotropy-driven microturbulence and magnetic-field evolution in
  shearing, collisionless plasma. {\em \mnras\/} {\bf 459}, 2701.

\bibitem[{Mikhailovskii} \& {Tsypin}(1971)]{Mikhailovskii:1971}
{\sc {Mikhailovskii}, A.~B. \& {Tsypin}, V.~S.} 1971 {Transport equations and
  gradient instabilities in a high pressure collisional plasma}. {\em Plasma
  Phys.\/} {\bf 13}, 785.

\bibitem[Pan \& Johnsen(2017)]{Pan:2017gt}
{\sc Pan, S. \& Johnsen, E.} 2017 {The role of bulk viscosity on the decay of
  compressible, homogeneous, isotropic turbulence}. {\em \jfm\/} {\bf 833},
  717.

\bibitem[{Perez} \& {Boldyrev}(2009)]{Perez2009}
{\sc {Perez}, J.~C. \& {Boldyrev}, S.} 2009 {Role of Cross-Helicity in
  Magnetohydrodynamic Turbulence}. {\em \prl\/} {\bf 102}, 025003.

\bibitem[Perez {\em et~al.\/}(2012)Perez, Mason, Boldyrev \&
  Cattaneo]{Perez:2012da}
{\sc Perez, J.~C., Mason, J., Boldyrev, S. \& Cattaneo, F.} 2012 On the energy
  spectrum of strong magnetohydrodynamic turbulence. {\em Phys.~Rev.~X\/} {\bf
  2}, 041005.

\bibitem[{Riley} {\em et~al.\/}(1996){Riley}, {Sonett}, {Tsurutani}, {Balogh},
  {Forsyth} \& {Hoogeveen}]{Riley1996}
{\sc {Riley}, P., {Sonett}, C.~P., {Tsurutani}, B.~T., {Balogh}, A., {Forsyth},
  R.~J. \& {Hoogeveen}, G.~W.} 1996 {Properties of arc-polarized Alfv{\'e}n
  waves in the ecliptic plane: Ulysses observations}. {\em \jgr\/} {\bf 101},
  19987.

\bibitem[Rosenbluth(1956)]{Rosenbluth:1956}
{\sc Rosenbluth, M.~N.} 1956 The stability of the pinch. {\em Los Alamos Sci.
  Lab. Rep.\/} {\bf LA-2030}.

\bibitem[Santos-Lima {\em et~al.\/}(2014)Santos-Lima, de~Gouveia Dal~Pino,
  Kowal, Falceta-Gon{\c c}alves, Lazarian \& Nakwacki]{SantosLima:2014cn}
{\sc Santos-Lima, R., de~Gouveia Dal~Pino, E.~M., Kowal, G., Falceta-Gon{\c
  c}alves, D., Lazarian, A. \& Nakwacki, M.~S.} 2014 Magnetic field
  amplification and evolution in turbulent collisionless magnetohydrodynamics:
  An application to the intracluster medium. {\em \apj\/} {\bf 781}, 84.

\bibitem[Schekochihin {\em et~al.\/}(2009)Schekochihin, Cowley, Dorland,
  Hammett, Howes, Quataert \& Tatsuno]{Schekochihin:2009eu}
{\sc Schekochihin, A.~A., Cowley, S.~C., Dorland, W., Hammett, G.~W., Howes,
  G.~G., Quataert, E. \& Tatsuno, T.} 2009 Astrophysical gyrokinetics: Kinetic
  and fluid turbulent cascades in magnetized weakly collisional plasmas. {\em
  \apjss\/} {\bf 182}, 310.

\bibitem[Schekochihin {\em et~al.\/}(2008)Schekochihin, Cowley, Kulsrud, Rosin
  \& Heinemann]{Schekochihin:2008en}
{\sc Schekochihin, A.~A., Cowley, S.~C., Kulsrud, R.~M., Rosin, M.~S. \&
  Heinemann, T.} 2008 Nonlinear growth of firehose and mirror fluctuations in
  astrophysical plasmas. {\em \prl\/} {\bf 100}, 081301.

\bibitem[Schekochihin {\em et~al.\/}(2010)Schekochihin, Cowley, Rincon \&
  Rosin]{Schekochihin:2010bv}
{\sc Schekochihin, A.~A., Cowley, S.~C., Rincon, F. \& Rosin, M.~S.} 2010
  Magnetofluid dynamics of magnetized cosmic plasma: firehose and gyrothermal
  instabilities. {\em \mnras\/} {\bf 405}, 291.

\bibitem[Sharma {\em et~al.\/}(2006)Sharma, Hammett, Quataert \&
  Stone]{Sharma:2006dh}
{\sc Sharma, P., Hammett, G.~W., Quataert, E. \& Stone, J.~M.} 2006 Shearing
  box simulations of the {MRI} in a collisionless plasma. {\em \apj\/} {\bf
  637}, 952.

\bibitem[Snyder {\em et~al.\/}(1997)Snyder, Hammett \& Dorland]{Snyder:1997fs}
{\sc Snyder, P.~B., Hammett, G.~W. \& Dorland, W.} 1997 {L}andau fluid models
  of collisionless magnetohydrodynamics. {\em \pop\/} {\bf 4}, 3974.

\bibitem[Squire {\em et~al.\/}(2017{\natexlab{{\em a\/}}})Squire, Kunz,
  Quataert \& Schekochihin]{Squire:2017ej}
{\sc Squire, J., Kunz, M.~W., Quataert, E. \& Schekochihin, A.~A.}
  2017{\natexlab{{\em a\/}}} Kinetic simulations of the interruption of
  large-amplitude shear-{A}lfv\'en waves in a high-$\ensuremath{\beta}$ plasma.
  {\em \prl\/} {\bf 119}, 155101.

\bibitem[Squire {\em et~al.\/}(2016)Squire, Quataert \&
  Schekochihin]{Squire:2016ev}
{\sc Squire, J., Quataert, E. \& Schekochihin, A.~A.} 2016 A stringent limit on
  the amplitude of {A}lfv{\'e}nic perturbations in high-beta low-collisionality
  plasmas. {\em \apjl\/} {\bf 830}, L25.

\bibitem[Squire {\em et~al.\/}(2017{\natexlab{{\em b\/}}})Squire, Schekochihin
  \& Quataert]{Squire:2016ev2}
{\sc Squire, J., Schekochihin, A.~A. \& Quataert, E.} 2017{\natexlab{{\em
  b\/}}} {Amplitude limits and nonlinear damping of shear-Alfv{\'e}n waves in
  high-beta low-collisionality plasmas}. {\em \njp\/} {\bf 19}, 055005.

\bibitem[Sulem \& Passot(2015)]{2015JPlPh..81a3203S}
{\sc Sulem, P.~L. \& Passot, T.} 2015 {{L}andau fluid closures with nonlinear
  large-scale finite Larmor radius corrections for collisionless plasmas}. {\em
  J. Plasma Phys.\/} {\bf 81}, 325810103.

\bibitem[{Tenerani} \& {Velli}(2018)]{Tenerani2018}
{\sc {Tenerani}, A. \& {Velli}, M.} 2018 Nonlinear firehose relaxation and
  constant-{B} field fluctuations. {\em \apj\/} {\bf 867}, L26.

\bibitem[{Tsurutani} {\em et~al.\/}(1994){Tsurutani}, {Ho}, {Smith},
  {Neugebauer}, {Goldstein}, {Mok}, {Arballo}, {Balogh}, {Southwood} \&
  {Feldman}]{Tsurutani1994}
{\sc {Tsurutani}, B.~T., {Ho}, C.~M., {Smith}, E.~J., {Neugebauer}, M.,
  {Goldstein}, B.~E., {Mok}, J.~S., {Arballo}, J.~K., {Balogh}, A.,
  {Southwood}, D.~J. \& {Feldman}, W.~C.} 1994 {The relationship between
  interplanetary discontinuities and Alfv{\'e}n waves: Ulysses observations}.
  {\em \grl\/} {\bf 21}, 2267.

\bibitem[{Tu} \& {Marsch}(1993)]{1993JGR....98.1257T}
{\sc {Tu}, C.-Y. \& {Marsch}, E.} 1993 {A model of solar wind fluctuations with
  two components - Alfven waves and convective structures}. {\em \jgr\/} {\bf
  98}, 1257.

\bibitem[{Vasquez} \& {Hollweg}(1998)]{Vasquez1998}
{\sc {Vasquez}, B.~J. \& {Hollweg}, J.~V.} 1998 {Formation of spherically
  polarized Alfv{\'e}n waves and imbedded rotational discontinuities from a
  small number of entirely oblique waves}. {\em \jgrsp\/} {\bf 103}, 335.

\bibitem[{Yang} {\em et~al.\/}(2017){Yang}, {Matthaeus}, {Parashar},
  {Haggerty}, {Roytershteyn}, {Daughton}, {Wan}, {Shi} \&
  {Chen}]{2017PhPl...24g2306Y}
{\sc {Yang}, Y., {Matthaeus}, W.~H., {Parashar}, T.~N., {Haggerty}, C.~C.,
  {Roytershteyn}, V., {Daughton}, W., {Wan}, M., {Shi}, Y. \& {Chen}, S.} 2017
  {Energy transfer, pressure tensor, and heating of kinetic plasma}. {\em
  \pop\/} {\bf 24}, 072306.

\bibitem[{Zhdankin} {\em et~al.\/}(2016){Zhdankin}, {Boldyrev} \&
  {Uzdensky}]{Zhdankin2016}
{\sc {Zhdankin}, V., {Boldyrev}, S. \& {Uzdensky}, D.~A.} 2016 {Scalings of
  intermittent structures in magnetohydrodynamic turbulence}. {\em \pop\/} {\bf
  23}, 055705.

\end{thebibliography}

\end{document}